\documentclass[aps,prd,twocolumn,showpacs,floatfix,preprintnumbers,amsmath,amssymb,nofootinbib,groupedaddress,superscriptaddress]{revtex4-1}

\input epsf
\usepackage{graphicx}
\usepackage{color}

\newcommand{\beq}{\begin{equation}}
\newcommand{\eeq}{\end{equation}}
\newcommand{\barr}{\begin{eqnarray}}
\newcommand{\earr}{\end{eqnarray}}

\newcommand{\rme}{\textrm{e}}

\newcommand{\rmd}{\textrm{d}}

\newcommand{\bs}{\boldsymbol}

\newcommand{\physrep}{Phys. Rep.}

\newcommand{\lsim}{\mathrel{\hbox{\rlap{\lower.55ex\hbox{$\sim$}} \kern-.3em \raise.4ex \hbox{$<$}}}}
\newcommand{\gsim}{\mathrel{\hbox{\rlap{\lower.55ex\hbox{$\sim$}} \kern-.3em \raise.4ex \hbox{$>$}}}}
\begin{document}
\title{Slowly-rotating stars and black holes in dynamical Chern-Simons gravity}

\author{Yacine Ali-Ha\"imoud} 
\email{yacine@ias.edu}
\affiliation{California Institute of Technology, Mail Code 350-17, Pasadena, California 91125}
\affiliation{Institute for Advanced Study, Einstein Drive, Princeton, New Jersey 08540}
\author{Yanbei Chen} 
\affiliation{California Institute of Technology, Mail Code 350-17, Pasadena, California 91125}

\date{\today}
\begin{abstract}

Chern-Simons (CS) modified gravity is an extension to general relativity (GR) in which the metric is coupled to a scalar field, resulting in modified Einstein field equations. In the dynamical theory, the scalar field is itself sourced by the Pontryagin density of the space-time. In this paper, the coupled system of equations for the metric and the scalar field is solved numerically for slowly-rotating neutron stars described with realistic equations of state and for slowly-rotating black holes. An analytic solution for a constant-density nonrelativistic object is also presented. It is shown that the black hole solution cannot be used to describe the exterior spacetime of a star as was previously assumed. In addition, whereas previous analysis were limited to the small-coupling regime, this paper considers arbitrarily large coupling strengths. It is found that the CS modification leads to two effects on the gravitomagnetic sector of the metric: $(i)$ Near the surface of a star or the horizon of a black hole, the magnitude of the gravitomagnetic potential is decreased and frame-dragging effects are reduced in comparison to GR. $(ii)$ In the case of a star, the angular momentum $J$, as measured by distant observers, is enhanced in CS gravity as compared to standard GR. For a large coupling strength, the near-zone frame-dragging effects become significantly screened, whereas the far-zone enhancement saturate at a maximum value $\Delta J_{\max} \sim (M/R) J_{\rm GR}$. Using measurements of frame-dragging effects around the Earth by Gravity Probe B and the LAGEOS satellites, a weak but robust constraint is set to the characteristic CS lengthscale, $\xi^{1/4} \lesssim 10^8$ km.
 
\end{abstract}

\maketitle
\section{Introduction}

Gravity, one of the four fundamental forces of nature, is elegantly described by Einstein's theory of general relativity (GR). Nearly a century after its discovery, GR has successfully passed the more and more subtle and precise tests that it has been submitted to (for a review, see for example Ref.~\cite{Will_06}). Nevertheless, Einstein's theory is probably not the final word on gravity. We expect that a more fundamental theory unifying all forces should be able to describe not only gravity but also quantum phenomena that may take place, for example, at the center of black holes. Since gravity does seem to describe nature quite faithfully at the energy and length scales that are accessible to us, we expect that it may be the low-energy limit of such a fundamental theory. If this is the case, gravity should be described by an effective theory, for which the action contains higher-order curvature terms than standard GR, the effect of which can become apparent in strong gravity situations.

One such theory is modified Chern-Simons (CS) gravity (for a review on the subject, see Ref.~\cite{Alexander_Yunes_09}). In this theory, the metric is coupled to a scalar field through the Pontryagin density $\bs{R \tilde{R}}$ (to be defined below). The dynamical and nondynamical versions of the theory (which in fact are two separate classes of theories), then differ in the prescription for the scalar field. In nondynamical CS gravity, the scalar field is assumed to be externally prescribed. It is often taken to be a linear function of coordinate time (the so-called ``canonical choice''), which selects a particular direction for the flow of time \cite{Jackiw_Pi_03}, and induces parity violation in the theory. Nondynamical CS gravity then depends on a single free parameter, which has been constrained with measurement of frame-dragging on bodies orbiting the Earth \cite{Smith_08}, and with the double-binary-pulsar \cite{Yunes_Spergel_09, Ali-Haimoud_11}. Nondynamical CS theory is quite contrived as a valid solution for the spacetime must satisfy the Pontryagin constraint $\bs{R \tilde{R}} = 0$; it should therefore rather be taken as a toy model used to gain some insight in parity-violating gravitational theories.

Dynamical CS gravity, which is the subject of the present paper, is a more natural theory where the scalar field itself is given dynamics (even though arbitrariness remains in the choice of the potential for the scalar field). The scalar field evolution equation is sourced by the Pontryagin density, which is non-vanishing only for spacetimes which are not reflection-invariant, as is the case in the vicinity of rotating bodies. Contrary to the non-dynamical theory, though, dynamical CS gravity is \emph{not} parity breaking, but simply has different solutions than GR for spacetimes which are not reflection-invariant (see discussion in Sec.~2.4 of Ref.~\cite{Alexander_Yunes_09}). Dynamical CS gravity has only recently received some attention, as it is more complex than the nondynamical version. Refs.~\cite{Yunes_Pretorius_09, Konno_09} computed the CS correction to the Kerr metric in the slow-rotation approximation. Ref.~\cite{Sopuerta_Yunes_09} studied the effect on CS gravity on the waveforms of extreme- and intermediate-mass ratio inspirals. Recently, Ref.~\cite{Yunes_et_al_10}, proposed a solution for the spacetime inside slowly-rotating neutron stars, assuming that the solution outside the star was identical to that of a black hole of the same mass and angular momentum. All the aforementioned studies were done in the \emph{small-coupling limit}, i.e.~considering the CS modification as a perturbation around standard GR.

The two main points of the present work are as follows. First, we show that, in contrast with GR, the spacetime around a slowly-rotating relativistic star is \emph{different} from that of a slowly-spinning black hole in CS gravity, owing to different boundary conditions (regularity conditions at the horizon for a black hole versus continuity and smoothness conditions at the surface of a star). Second, we solve for the CS modification to the metric and the scalar field simultaneously, in the fully-coupled case (nonperturbative with respect to the CS coupling strength), for a slowly-rotating star or black hole. Our motivation in doing so is that frame-dragging effects are difficult to measure and are not highly constrained; it is therefore still possible that they differ significantly from the GR prescription. The solutions obtained would be exact (modulo the slow-rotation approximation) if CS gravity were taken as an exact theory. If one asumes the CS action is only the truncated series expansion of an exact theory, then our solutions are only meaningful at the linear order in the CS coupling strength.

As shown in previous works, the CS correction only affects the gravitomagnetic sector of the metric at leading order in the slow-rotation limit. We find that for both stars and black holes, the CS correction leads to a \emph{suppression} of frame-dragging at a distance of a few stellar radii, or a few times the black hole horizon radius. This suppression is perturbative in the small-coupling regime but can become arbitrarily large in the nonlinear regime, asymptotically leading to a complete screening of frame-dragging effects near the star or black hole in the large-coupling-strength limit. Far from the boundary of a star, the magnitude of frame-dragging effects is \emph{enhanced}: we find a correction to the $t \phi$ metric component $\Delta g^{t \phi} \approx -2 \Delta J_{\rm CS}/r^3$ at large radii, with $\Delta J_{\rm CS} > 0$. This means that for a given angular rotation rate $\Omega$, the angular momentum measured by distant observers is enhanced as compared to standard GR. We evaluate the correction to the angular momentum $\Delta J_{\rm CS} \equiv J_{\rm CS} - J_{\rm GR}$ as a function of the coupling strength and find that it increases quadratically with the coupling parameter in the small-coupling regime, and asymptotes to a constant value $\Delta J_{\max} \sim (M/R) J_{\rm GR}$ in the large-coupling regime.

These results are obtained using an analytic approximation for constant-density nonrelativistic objects, and confirmed with a numerical solution for neutron stars described with realistic equations of state. The black hole solution is computed numerically and checked against known analytic solutions in the small coupling regime.

Finally, using measurements of frame-dragging effects around the Earth, we set a weak but robust constraint to the CS characteristic lengthscale, $\xi^{1/4} \lesssim 10^8$ km. We argue that this bound is the only current astrophysical constraint to the theory.

This paper is organized as follows: in Sec.~\ref{sec:basics}, we review the theory of dynamical Chern-Simons gravity and define our notation. In Sec.~\ref{sec:slowrot}, we lay out the general formalism to compute the CS scalar field and the metric in CS gravity, for a slowly rotating object. We provide analytic expressions for the exterior solution in some limiting cases in Sec.~\ref{sec:analytic}, as a well as an analytic solution for the full spacetime for nonrelativistic constant-density stars.
We present the results of our numerical computations for realistic neutron stars in Sec.~\ref{sec:numerical} and for black holes in Sec.~\ref{sec:black hole}. We discuss constraints to the theory in Sec.~\ref{sec:constraints} and conclude in Sec.~\ref{sec:conclusion}. 

Throughout this paper we use geometric units $G = c = 1$. We adopt the conventions of Ref.~\cite{MTW} for the signature of the metric, Riemann and Einstein tensors.

\section{Dynamical Chern-Simons gravity} \label{sec:basics}

For a review on Chern-Simons modified gravity, we refer the reader to Ref.~\cite{Alexander_Yunes_09}. Here we simply recall the main equations and results and define our notation.

We consider the following action defining the modified theory \footnote{The conversion from the notation used in Ref.~\cite{Alexander_Yunes_09} (AY09) to that of the present work is given by $\vartheta \equiv \sqrt{\beta} \vartheta_{\rm AY09}$, $16 \pi \ell^4_{\rm cs} \equiv \alpha^2/(\kappa \beta) \equiv \xi$, $V(\vartheta) \equiv \beta V_{\rm AY09}(\vartheta/\sqrt{\beta})$. The conversion from the notation of Ref.~\cite{Smith_08} (who define the Riemann tensor with an opposite sign) is given by $\ell^2_{\rm cs} = \ell/3$.}:
\barr
S &=& \int d^4 x \sqrt{- g} \left[\frac{1}{16 \pi} R + \mathcal{L}_{\rm mat}\right]\nonumber\\
&+& \int d^4 x \sqrt{- g} \left[\frac{\ell_{\rm cs}^2}{4} \vartheta \boldsymbol{R \tilde{R}} - \frac12 \nabla_{\mu} \vartheta \nabla^{\mu} \vartheta  - V(\vartheta) \right]. \label{eq:action}
\earr
In Eq.~(\ref{eq:action}), the first term contains the standard Einstein-Hilbert action and the matter contribution with lagrangian density $\mathcal{L}_{\rm mat}$. The second term is the CS modification, which only depends on the dimensionfull coupling constant $\ell_{\rm cs}$ (which has dimensions of length; its relation to the parameter $\xi$ of Ref.~\cite{Yunes_Pretorius_09} is $\xi \equiv 16 \pi \ell_{\rm cs}^4$), and of course on the shape of the potential $V$ for the dimensionless CS scalar field $\vartheta$. The Pontryagin density $\boldsymbol{R\tilde{R}}$ is given by the contraction of the Riemann tensor $R_{\alpha \beta \mu \nu}$ with its dual:
\beq
\boldsymbol{R\tilde{R}} \equiv \frac12 \epsilon^{\mu \nu \sigma \tau}R^{\alpha \beta}_{~~\sigma \tau}R_{\beta \alpha \mu \nu},
\eeq
where $\epsilon^{\mu \nu \sigma \tau}$ is the four-dimensional Levi-Civita tensor (with convention $\epsilon^{0123} = +1$ in a right-handed orthonormal basis). The equations of motion resulting from the modified action are the following:

\emph{(i)} The modified Einstein field equations:
\beq
G_{\mu \nu} + 16 \pi \ell^2_{\rm cs} C_{\mu \nu} = 8 \pi \left(T_{\mu \nu}^{\rm mat} + T_{\mu \nu}^{\vartheta}\right), \label{eq:EF}
\eeq
where $G_{\mu \nu}$ is the Einstein tensor, $T_{\mu \nu}^{\rm mat}$ is the matter stress-energy tensor,
\beq
C^{\mu \nu} \equiv \partial_{\sigma} \vartheta ~\epsilon^{\sigma \alpha \beta (\mu } \nabla_{\alpha} R^{\nu)}_{~\beta} + \frac12 \nabla_{\tau}(\partial_{\sigma} \vartheta) ~\epsilon^{\alpha \beta \sigma (\mu} R^{\nu) \tau}_{~~~\beta \alpha}
\eeq
is a four-dimensional generalization of the Cotton-York tensor, and 
\beq
T_{\mu \nu}^{\vartheta} \equiv \nabla_{\mu} \vartheta \nabla_{\nu}\vartheta - \frac12 g_{\mu\nu} \nabla_{\alpha}\vartheta \nabla^{\alpha}\vartheta - g_{\mu \nu} V(\vartheta) \label{eq:Tmunu.theta}
\eeq
is the stress-energy tensor associated with the scalar field $\vartheta$.

\emph{(ii)} The evolution equation for the scalar field $\vartheta$:
\beq
\square \vartheta =  - \frac{\ell^2_{\rm cs}}{4} \boldsymbol{R \tilde{R}} + V'(\vartheta), \label{eq:theta-evol}
\eeq
where $\square \equiv g^{\mu \nu} \nabla_{\mu}\nabla_{\nu}$ is the usual covariant d'Alembertian operator.

In general, the fundamental theory from which CS gravity arises should predict a shape for the potential $V(\vartheta)$. This would introduce an additional free parameter (or several parameters) in addition to $\ell_{\rm cs}^2$ . Since our goal here is mainly to study the effect of the coupling strength $\ell_{\rm cs}^2$, we follow previous works and assume $V(\vartheta) = 0$ for simplicity.

\section{Slowly-rotating relativistic stars in Chern-Simons gravity}\label{sec:slowrot}

\subsection{Metric}

We consider a stationary, axially symmetric system. This means that there exist a time coordinate $x^0 = t$ and an angular coordinate $x^3 = \phi$ such that the metric components do not depend on $t$ nor $\phi$; if $x^1$ and $x^2$ are the two remaining spatial coordinates, we have
\beq
g_{\mu \nu} = g_{\mu \nu}(x^1, x^2). \label{eq:Killing}
\eeq
We moreover \emph{assume} that the line element $ds^2 = g_{\mu \nu} dx^{\mu} d x^{\nu}$ satisfies an additional discrete symmetry: we suppose that the effect of inversion of the azimuthal angle is identical to that of inverting time, $\phi \rightarrow - \phi \Leftrightarrow t \rightarrow - t$, i.e. 
\barr
&&ds^2(t,x^1, x^2, -\phi; dt, dx^1, dx^2, -d\phi)\\ 
= ~&&ds^2(-t,x^1, x^2, \phi; -dt, dx^1, dx^2,d\phi). \label{eq:assumption}
\earr
Note that one could clearly not have made this assumption in nondynamical CS gravity where the externally prescribed scalar field defines the flow of time (for the ``canonical'' choice of scalar field). The validity of this assumption in dynamical CS gravity will be justified \emph{a posteriori} by checking that solutions do exist with this discrete symmetry (however, it does not guarantee the uniqueness of the solutions found). With this assumption, we find that\footnote{In GR, one does not need the additional assumption (\ref{eq:assumption}) to obtain (\ref{eq:off-diag}), see proof in Section 7.1 of Ref.~\cite{Wald}. The proof of Ref.~\cite{Wald} makes use of the GR Einstein field equation, however, and can therefore not be carried over to CS gravity. One could also retain the a priori non-zero $g_{tr}$ and $g_{t \theta}$ metric components and show that the CS field equations lead them to vanish. We have not chosen this path in order ot avoid excessive algebra.} 
\beq
g_{t x^1} = g_{t x^2} = g_{\phi x^1} = g_{\phi x^2}= 0. \label{eq:off-diag}
\eeq

By an appropriate choice of coordinate transformations of the form $\tilde{x}^{i} = f^{i}(x^1, x^2)$ for $i = 1, 2$, one may rewrite the line element in the form \cite{Hartle_Sharp_67}
\barr
ds^2 = &-& \rme^{2 \Phi(r, \theta)} dt^2 + \rme^{2 \Lambda(r, \theta)} dr^2 \nonumber\\
&+& r^2 H(r, \theta)\left(d \theta^2 + (\sin \theta)^2 d \phi^2\right)\nonumber\\
 &-& 2 r^2 (\sin \theta)^2 \omega(r, \theta) dt d\phi, \label{eq:metric}
\earr
where we have removed the tildes for clarity and denoted $x^1 = r$, $x^2 = \theta$.

\subsection{Stress-energy tensor}

We consider an ideal fluid in solid rotation with angular rate $d \phi/dt = \Omega$, i.e. such that the 4-velocity is of the form
\beq
u^{\mu} = \{u^t, u^r, u^{\theta}, u^{\phi}\} = u^0(r, \theta)\{1, 0, 0, \Omega\}.
\eeq
To evaluate $u^0$, we use the normalization condition $g_{\mu \nu} u^{\mu} u^{\nu} = -1$, i.e. 
\beq
u^0= \left(\rme^{2 \Phi} + 2 r^2 (\sin \theta)^2 \omega \Omega + r^2 H (\sin \theta)^2 \Omega^2\right)^{-1/2}.
\eeq
The stress-energy tensor of a perfect fluid is then obtained from
\beq
T^{\mu \nu} = (\rho + p) u^{\mu} u^{\nu} + p g^{\mu \nu},
\eeq
where $\rho$ and $p$ are the density and pressure of the fluid as measured in its local rest frame. 

\subsection{Slow-rotation expansion}

We consider a star\footnote{In this paper we use the word `star' to refer to a general (non-empty) astrophysical object.} of radius $R$ in solid rotation with angular velocity $\Omega$. The star is in \emph{slow rotation} if $\Omega R \ll 1$. In what follows we shall expand the metric functions $\Phi, \Lambda, H, \omega, g_{tr}$ and $g_{t \theta}$ as well as the density and pressure fields in powers of $\Omega R$. 

Under the change of coordinates $\phi' = - \phi$, the angular velocity changes sign, $\Omega' = -\Omega$, $g_{tt}, g_{rr}, g_{\theta\theta}, g_{\phi \phi}$ and therefore $\Phi, \Lambda, H$ remain unchanged, while $g_{t \phi}$ and therefore $\omega$ change sign. Similarly, considerations of the transformation properties of the stress-energy tensor show that $\rho$ and $p$ remain unchanged. Since the field equation expressed in the $t, r, \theta, \phi'$ coordinate system are the same as those in the $t, r, \theta, \phi$ coordinate system for a fluid with angular velocity $-\Omega$ (because there is no functional dependence on $\phi$), we conclude that $\Phi, \Lambda, H, \rho$ and $p$ are even functions of $\Omega$ whereas $\omega$ is an odd function of $\Omega$. 

In the slow-rotation approximation, we only keep the lowest-order contribution to each function. From the previous discussion, this means that we keep only the $\mathcal{O}(1)$ terms in $\Phi, \Lambda, H, \rho$ and $p$ and the $\mathcal{O}(\Omega R)$ term in $\omega$. The next-order contributions are of relative order $f(\ell_{\rm cs}^2/M^2)\times (\Omega R)^2$ in CS gravity, where $f$ is an unknown function of the coupling strength. It is beyond the scope of this paper to evaluate $f$, but we emphasize that our slow-rotation approximation is only valid as long as $f \times (\Omega R)^2 \ll 1$.

To lowest order, the velocity of the fluid becomes
\beq
u^{\mu} \approx \rme^{- 2 \Phi}\{1, 0, 0, \Omega\}.
\eeq
Moreover, the Pontryagin density is of order $\mathcal{O}(\Omega R)$ (we give the exact expression below). Therefore, the scalar field $\vartheta$ is also of order $\mathcal{O}(\Omega R)$ [from Eq.(\ref{eq:theta-evol}) with $V=0$]. As a consequence, the components of the scalar field stress-energy tensor $T_{\mu \nu}^{\vartheta}$ [Eq.~(\ref{eq:Tmunu.theta})] are of order $\mathcal{O}(\Omega R)^2$ and need not be considered at this level of approximation, i.e. the energy of the scalar field does not curve space-time to first order in the rotation rate.

\subsection{Stellar structure equations for $\Phi, \Lambda$ and $H$}

The leading-order contribution to the metric components which are even in $\Omega$ can be obtained in the non-rotating case, $\Omega = 0$. In that case the system is spherically symmetric and the Pontryagin density vanishes. The scalar field being unsourced, the Einstein field equations are identical to those of standard GR. By spherical symmetry, $\Phi, \Lambda, H, \rho$ and $p$ are functions of $r$ only. By rescaling the radial coordinate, one can moreover set $H$ to unity. The functions $\Phi(r)$ and $\Lambda(r)$ then satisfy the usual relativistic stellar structure equations \cite{MTW}:
\beq
\Lambda(r) \equiv -\frac12 \ln\left(1 - \frac{2 m(r)}{r}\right),
\eeq
with
\beq
m(r) \equiv \int_0^r 4 \pi r'^2 \rho(r') dr'
\eeq
being the mass enclosed within a radius $r$. Inside the star, the function $\Phi(r)$ is the solution of the equation
\beq
\Phi'(r) = \frac{m + 4 \pi r^3 p}{r(r - 2m)},
\eeq
where the pressure $p(r)$ must satisfy the equation of hydrostatic equilibrium
\beq
p'(r) = -(\rho + p) \frac{m + 4 \pi r^3 p}{r(r - 2m)}.
\eeq
Outside the star, we have $p = \rho = 0$, and 
\beq
\Phi(r) = - \Lambda(r) = \frac12 \ln\left(1 - \frac{2M}{r}\right),
\eeq
where $M$ is the total mass of the star. 

\subsection{Gravitomagnetic sector}

\subsubsection{General equations}

Let us now consider the $t\phi$ component of the metric. Evaluating the $t\phi$ component of the modified field equation to lowest order in $\Omega R$, we find, in agreement with Ref.~\cite{Yunes_et_al_10} and Refs.~\cite{Yunes_Pretorius_09, Konno_09} outside the star:
\barr
&&\left(1 - \frac{2 m}{r}\right) \frac{\partial^2 \omega}{\partial r^2} + \frac{4}{r}\left(1 - \frac{2 m}{r} - \pi (\rho + p) r^2\right) \frac{\partial \omega}{\partial r} \nonumber\\
&& + \frac1{r^2}\left(\frac{\partial^2 \omega}{\partial \theta^2} + 3 \textrm{cotan} \theta \frac{\partial \omega}{\partial \theta}\right) + 16 \pi (\rho + p)(\Omega - \omega) \nonumber\\
&&= -96 \pi \ell_{\rm cs}^2 \frac{\rme^{\Phi - \Lambda}}{r^4 \sin \theta} \frac{\partial}{\partial r}\Bigg{[} \left(\frac{m}{r} - \frac{4 \pi}3 \rho r^2 \right)\frac{\partial \vartheta}{\partial \theta}\Bigg{]}. \label{eq:ODE-omega}
\earr
In addition, the evolution equation for the CS scalar field becomes, in the time-independent case and to first order in $\Omega R$:
\barr
&&\left(1 - \frac{2 m}{r}\right) \frac{\partial^2\vartheta}{\partial r^2} + \frac{2}{r}\left(1 - \frac{m}{r} + 2 \pi (p - \rho) r^2\right) \frac{\partial \vartheta}{\partial r} \nonumber\\
&&+ \frac{1}{r^2}\left(\frac{\partial^2 \vartheta}{\partial \theta^2} + \textrm{cot} \theta \frac{\partial \vartheta}{\partial \theta} \right) = - \frac{\ell^2_{\rm cs}}4 \bs{R \tilde R} \label{eq:ODE-Theta},
\earr
where, to lowest order, the Pontryagin density is given by
\beq
\bs{R \tilde R}  = -24 \frac{\rme^{-(\Phi + \Lambda)}}{\sin \theta}\left(\frac{m}{r^3} - \frac{4 \pi}{3} \rho\right) \frac{\partial}{\partial \theta}\left(\sin^2 \theta \frac{\partial \omega}{\partial r} \right).
\eeq
Finally, we also note that all other non-diagonal components of the Einstein, ``C'' and stress-energy tensors vanish. Therefore the metric (\ref{eq:metric}) is indeed a valid solution of the modified Einstein field equations provided we solve for $\Phi$, $\Lambda$ and $\omega$ as described above.

\subsubsection{Multipole expansion}

Following Ref.~\cite{Konno_09}, we decompose $\vartheta(r, \theta)$ on the basis of Legendre polynomials and $\omega(r, \theta)$ on the basis of their derivatives:
\barr
\vartheta(r, \theta) &=& \sum_{l = 0}^{\infty} \vartheta_l(r) P_l(\cos \theta),\\
\omega(r, \theta) &=& \sum_{l=1}^{\infty} \omega_l(r) P_l'(\cos \theta).
\earr
Eqs.~(\ref{eq:ODE-omega}) and (\ref{eq:ODE-Theta}) can then be rewritten as an infinite set of coupled differential equations for the coefficients $\omega_l$ and $\vartheta_l$. Each pair of equations is independent in the sense that the $l$-th multipole of $\omega$ only couples to the $l$-th multipole of $\vartheta$ \cite{Konno_09}. The only sourced multipole is $l=1$, and therefore $\vartheta_l = 0$ and $\omega_l = 0$ for $l \neq 1$ (this follows from the system being second order in $\omega_l$ and $\vartheta_l$, and the requirement that $\vartheta$ and $\omega$ vanish at large radii and be bounded at the stellar center). The functions $\vartheta_1(r)$ and $\omega_1(r) = \omega(r)$ are the solutions of the following system
\barr
&&\left(1 - \frac{2 m}{r}\right) \omega'' + \frac{4}{r}\left(1 - \frac{2 m}{r} - \pi (\rho + p) r^2\right) \omega' \nonumber\\
&& + 16 \pi (\rho + p)(\Omega - \omega) \nonumber\\
&&= 96 \pi \frac{\ell_{\rm cs}^2}{r^4} \rme^{\Phi - \Lambda} \frac{d}{dr}\Bigg{[}\left(\frac{m}{r} - \frac{4\pi}3 \rho r^2\right) \vartheta_1 \Bigg{]}~~~~~\label{eq:ODE-omega1}
\earr
and
\barr
&&\left(1 - \frac{2 m}{r}\right) \vartheta_1'' + \frac{2}{r}\left(1 - \frac{m}{r} + 2 \pi (p - \rho) r^2\right) \vartheta_1' - \frac{2}{r^2} \vartheta_1\nonumber\\
&& = 12 \ell^2_{\rm cs} \rme^{-(\Phi + \Lambda)}\left(\frac{m}{r^3} - \frac{4 \pi}{3} \rho\right)\omega'. \label{eq:ODE-theta1}
\earr
One can decompose $\omega$ into two pieces: a part that would be present in standard GR, $\omega_{\rm GR}$, that can be obtained by setting $\ell_{\rm cs}^2 = 0$ in Eq.~(\ref{eq:ODE-omega1}), and a correction (not necessarily small) $\Delta \omega_{\rm CS} \equiv \omega - \omega_{\rm GR}$, which is the solution of Eq.~(\ref{eq:ODE-omega1}) with $\Omega = 0$. Note that it is the full $\omega = \omega_{\rm GR} + \Delta \omega_{\rm CS}$ that sources the CS scalar field in Eq.~(\ref{eq:ODE-theta1}).

Before proceeding further, we shall discuss boundary conditions at the surface of the star. First, the function $\omega$ must be continuous at the boundary of the star. In addition, integrating Eq.~(\ref{eq:ODE-omega1}) between $R-\epsilon$ and $R+ \epsilon$ gives us a jump condition for the derivative of $\omega$ at the stellar surface:
\beq
\omega'(R+\epsilon) - \omega'(R - \epsilon) = \frac{96 \pi \ell_{\rm cs}^2}{R^2} \frac{4 \pi}{3} \rho(R-\epsilon) \vartheta_1(R). \label{eq:jump}
\eeq
For neutron stars, the surface density is nearly vanishing [in the sense that $\rho(R-\epsilon) \ll \rho(0)$], and $\omega'$ is (nearly) continuous at the surface. For constant density objects, however, there is a jump in the derivative of $\omega$ at the surface. Finally, inspection of Eq.~(\ref{eq:ODE-theta1}) shows that all coefficients are bounded (although potentially discontinuous), and therefore both $\vartheta_1$ and $\vartheta_1'$ must be continuous at the stellar surface, in all cases.

\subsubsection{Simplification outside the star} \label{sec:solution-outside}

Outside the star, $\rho = p = 0$ and Eq.~(\ref{eq:ODE-omega1}) can be integrated once and simplified to
\beq
\omega' = \frac{96 \pi \ell_{\rm cs}^2 M}{r^5} \vartheta_1 - \frac{6 J}{r^4}, \label{eq:omega'-out}
\eeq
where $J$ is a constant of integration. $J$ is also the total angular momentum of the star as measured by observers in the asymptotically flat far zone, as we shall discuss in Sec.~\ref{sec:note inertia}. We write $J = J_{\rm GR} + \Delta J_{\rm CS}$, where $J_{\rm GR}$ is the value of $J$ in standard GR and $\Delta J_{\rm CS}$ is the correction (not necessarily perturbative) that arises in CS gravity.

Using Eq.~(\ref{eq:omega'-out}) into Eq.~(\ref{eq:ODE-theta1}), we obtain an equation for $\vartheta_1$ alone:
\barr
\left(1 - \frac{2 M}{r}\right) \vartheta_1'' + \frac{2}{r}\left(1 - \frac{M}{r} \right) \vartheta_1'  - \left(\frac{2}{r^2} + \frac{9 \zeta^2 R^6}{r^8}\right)\vartheta_1\nonumber\\
= -72 \ell^2_{\rm cs} \frac{M J}{r^7},~~~~~ \label{eq:ODE-theta-out}
\earr 
where we have defined the dimensionless coupling strength $\zeta$, to be used repeatedly in the remainder of this paper:
\beq
\zeta^2 \equiv 128 \pi \frac{\ell_{\rm cs}^4 M^2}{R^6}. \label{eq:zeta}
\eeq
Physically, $\zeta^{1/2}$ is of the order of the ratio of the CS lengthscale to the dynamical timescale of the system. It is important to notice that the definition of $\zeta$ depends on the system considered through its average density.

Equation (\ref{eq:ODE-theta-out}) does not have an analytic solution in the general case, but it does have one in the two limiting cases of small coupling and nonrelativistic stars, discussed in Sec.~\ref{sec:analytic}.

Equations (\ref{eq:omega'-out}) and (\ref{eq:ODE-theta-out}) are also valid for a slowly-rotating black hole. Because a black hole does not have a surface of discontinuity but has a horizon, the boundary conditions for the scalar field and the metric are different than for a star. We shall discuss the black hole solution in Sec.~\ref{sec:black hole}.

\subsection{A note on the angular momentum and moment of inertia} \label{sec:note inertia}

There are several possible definitions for angular momentum in GR, and we therefore specify the definition that we use here. 

The angular momentum $\boldsymbol{J}$ of a body can be defined from the asymptotic behavior of the metric far outside the source \cite{MTW, Prior}:
\barr
\rmd s^2 &=& - \left[1 - \frac{2M}{r}\ + \mathcal{O}\left(\frac{1}{r^2}\right)\right] \rmd t^2 \nonumber\\
&&- \left[4 \frac{\epsilon_{ijk} J^j x^k}{r^3} + \mathcal{O} \left(\frac{1}{r^3}\right)\right]\rmd t \rmd x^i \nonumber\\
&&+ \left[1 + \frac{2M}{r}\ + \mathcal{O}\left(\frac{1}{r^2}\right)\right] \delta_{ij} \rmd x^i \rmd x^j \nonumber\\
&&+ \mathcal{O}\left(\frac{1}{r}\right) [1 - \delta_{ij}] \rmd x^i \rmd x^j.  
\earr
In spherical polar coordinates, and using the notation of Eq.~(\ref{eq:metric}), this corresponds to 
\beq
\omega = \frac{2J}{r^3} \left[1 + \mathcal{O}(1/r)\right].
\eeq
We therefore see that the constant of integration $J$ in Eq.~(\ref{eq:omega'-out}) corresponds to the total angular momentum of the star or black hole as measured by distant observers.

The moment of inertia is an ill-defined quantity for relativistic systems, as in general the angular momentum of a body in solid rotation may not scale linearly with the angular velocity. However, in the slow-rotation approximation, the angular momentum does scale linearly with $\Omega$, to first order. We can therefore \emph{define} the relativistic generalization of the Newtonian moment of inertia by \cite{Hartle_67}
\beq
I\equiv \lim_{R \Omega \rightarrow 0} \frac{J}{\Omega}.
\eeq
In general the moment of inertia will depend on the mass and equation of state of the considered object. In Chern-Simons gravity, it will also depend on the coupling constant $\ell_{\rm cs}^2$. 

In GR, there exists a simple integral formula for the moment of inertia \cite{Hartle_67}:
\beq
I_{\rm GR} = \frac{8 \pi}{3} \int_0^R r^4 (\rho + p) \rme^{\Lambda-\Phi} \left( 1 - \frac{\omega}{\Omega}\right) \rmd r.  \label{eq:IGR}
\eeq
This formula is \emph{a priori} valid \emph{only} in GR, and does not necessarily hold in modified gravity theories. Equation (\ref{eq:IGR}) is indeed a local definition, whereas we have defined angular momentum from its imprint on the spacetime in the far-field.

It turns out, however, that Eq.~(\ref{eq:IGR}) still holds in CS gravity, due to the particular form of Eq.~(\ref{eq:ODE-omega1}). To see this, let us first rewrite Eq.~(\ref{eq:ODE-omega1}) in the form
\barr
&&\frac1{r^4}\frac{d}{dr}\left(r^4 \rme^{-(\Phi + \Lambda)} \omega'\right) + 16 \pi (\rho + p) \rme^{\Lambda-\Phi}  \left(\Omega - \omega\right) \nonumber\\
&&=  96 \pi \frac{\ell_{\rm cs}^2}{r^4} \frac{d}{dr}\Bigg{[}\left(\frac{m}{r} - \frac{4\pi}3 \rho r^2\right) \vartheta_1 \Bigg{]}, \label{eq:omega_Hartle}
\earr
Multiplying Eq.~(\ref{eq:omega_Hartle}) by $r^4$, integrating form 0 to $R-\epsilon$, and using the jump condition for $\omega'$ at the surface of the star, Eq.~(\ref{eq:jump}), in conjunction with Eq.~(\ref{eq:omega'-out}), we recover Eq.~(\ref{eq:IGR}). We emphasize that this integral formula is not a definition of the moment of inertia and would not necessarily be valid in other modified gravity theories.


\section{Analytic approximate solutions} \label{sec:analytic}

Before tackling the full numerical solution of the problem, we give a few analytic results in some simple cases. 

\subsection{Analytic exterior solution in the small coupling limit}\label{sec:analytic-small-coupling}

If $\zeta \ll 1$, the differential equation for $\vartheta_1$ outside the star Eq.~(\ref{eq:ODE-theta-out}) has an analytic solution \cite{Yunes_Pretorius_09, Konno_09}, valid up to corrections of relative order $\mathcal{O}(\zeta^2)$:
\barr
&&\vartheta_1(\zeta \ll 1, r\geq R) = \frac58 \frac{J}{M^2} \frac{\ell^2_{\rm cs}}{M^2}\Bigg{\{}\frac{M^2}{r^2} + \frac{2M^3}{r^3} + \frac{18}{5}\frac{M^4}{r^4}\nonumber\\
&&+ C_2 (r - M) + 3 C_1\left[1 + \left(\frac{r}{2M} - \frac12\right) \ln(1 - 2M/r)\right] \Bigg{\}},~~~~~ \label{eq:outside-sol}
\earr
where $C_1$ and $C_2$ are constants of integration. The requirement that $\vartheta$ remains finite at infinity implies that $C_2 = 0$. In Refs.~\cite{Yunes_Pretorius_09} and \cite{Konno_09} where a rotating black hole was studied, $C_1$ was also (rightly) set to zero so $\vartheta$ (as well as $\omega$) remains finite at the horizon. In our case, however, since $r > 2M$ outside the star, the homogeneous solution proportional to $C_1$ is well behaved everywhere outside the star, and $C_1 \neq 0$ a priori (in fact, we shall show that $C_1 \approx 1$ for nonrelativistic objects). The integration constant $C_1$ must be  determined from the continuity and smoothness requirements for $\vartheta$ at the stellar boundary. This shows that the black hole solution cannot be used as the solution outside a star as was assumed in Ref.~\cite{Yunes_et_al_10}.

From Eq.~(\ref{eq:omega'-out}), we then obtain $\omega$ outside the star (setting the additional integration constant to zero so that $\omega(+\infty) = 0$):
\barr
&&\omega(\zeta \ll 1, r\geq R) = \frac{2 J}{r^3} + 10 \pi \frac{J \ell^4_{\rm cs}}{M^7} \Bigg{\{} - \frac{M^6}{r^6} - \frac{12}7 \frac{M^7}{r^7} \nonumber\\
&&- \frac{27}{10} \frac{M^8}{r^8}  + C_1 \Bigg{[} \frac{15}{32} \frac Mr + \frac{15}{32} \frac{M^2}{r^2} + \frac58 \frac{M^3}{r^3} - \frac{81}{16} \frac{M^4}{r^4}\nonumber\\
&&~~~~~~ + \ln\left(1 - 2M/r\right) \left(\frac{15}{64} - 3 \frac{M^3}{r^3} + \frac94 \frac{M^4}{r^4}\right) \Bigg{]} \Bigg{\}}. \label{eq:omega outside}
\earr
It will be useful in what follows to write the asymptotic behavior of $\vartheta$ and $\omega$ at large radii $r \gg M$ up to corrections of relative order $M/r$ (\emph{a fortiori}, these results are also valid everywhere outside the source in the nonrelativistic limit $R \gg M$): 
\barr
\vartheta_1(\zeta \ll1, r\gg M) &\approx& \frac58 \frac{J}{M^2} \frac{\ell^2_{\rm cs}}{M^2} \left[(1 - C_1)\frac{M^2}{r^2} -\frac{32}5 C_1 \frac{M^5}{r^5} \right],~~~~~ \label{eq:theta-expansion} \\
\omega(\zeta \ll 1, r \gg M) &\approx& \frac{2 J}{r^3} \nonumber\\
&+& 10 \pi \frac{J \ell^4_{\rm cs}}{M^7} \left[  (C_1 - 1) \frac{M^6}{r^6} + \frac{64}{15} C_1 \frac{M^9}{r^9} \right].~ \label{eq:domega far}
\earr
Before going further, let us assess the differences of our results with those of Ref.~\cite{Yunes_et_al_10}, who also worked in the small coupling regime. First, as we pointed out previously, Ref.~\cite{Yunes_et_al_10} set $C_1 = 0$ whereas we shall show below that $C_1 = 1 + \mathcal{O}(M/R) + \mathcal{O}(\zeta^2)$. Therefore our solution for the scalar field is of order $\mathcal{O}(M/R)$ smaller than that of Ref.~\cite{Yunes_et_al_10} at large distances. Second, and more importantly, in Ref.~\cite{Yunes_et_al_10}, the parameter $J$ used in Eq.~(\ref{eq:omega outside}) was set to $J_{\rm GR}$. In reality, $J = J_{\rm GR} + \Delta J_{\rm CS}$ is determined by imposing the continuity and jump conditions for $\omega$ at the surface of the star. In the small coupling regime, 
\beq
\Delta J_{\rm CS} \sim \zeta J_{\rm GR} \label{eq:DJ-general}
\eeq
and $\Delta J_{\rm CS}$ is in general nonzero. We therefore have $\Delta \omega_{\rm CS} \sim J \zeta/r^3$ at large distances, instead of $\Delta \omega_{\rm CS}\sim (R^6/M^3) J \zeta/r^6$ in Ref.~\cite{Yunes_et_al_10}. Physically, this means that the CS correction translates into a change of angular momentum as measured by distant observers, whereas the previous solution did not, strictly speaking, lead to any additional angular momentum.

\subsection{Analytic exterior solution in the nonrelativistic limit} \label{sec:analytic-zeta}

If $M \ll R$ then the solution of Eq.~(\ref{eq:ODE-theta-out}) is 
\barr
\vartheta_1(r \geq R \gg M) = \frac{J}{\sqrt{2 \pi} \zeta R^3} r \Big{[}1 - D_2 \cosh\left(\frac{ \zeta R^3}{r^3}\right)\nonumber\\
 + D_1 \sinh\left(\frac{\zeta R^3}{r^3}\right) \Big{]}, ~~~\label{eq:theta1-out-nonrel}
\earr
where $D_1$ and $D_2$ are integration constants. Requiring $\vartheta$ to be finite at large radii implies $D_2 = 1$, whereas $D_1$ needs to be fixed using continuity conditions at the stellar boundary.

From Eq.~(\ref{eq:omega'-out}), we then obtain $\omega$ outside the star (choosing the additional integration constant such that $\omega(+\infty) = 0$):
\barr
\omega(r \geq R \gg M) = \frac{2 J}{R^3 \zeta}\Bigg{\{}\sinh\left(\frac{ \zeta R^3}{r^3}\right)\nonumber\\
 + D_1 \left[1 - \cosh\left(\frac{\zeta R^3}{r^3}\right)\right] \Bigg{\}}.\label{eq:omega-out-zeta}
\earr
If we Taylor-expand these solutions for $\zeta \ll1$, we should recover the analytic solution obtained in Sec.~\ref{sec:analytic-small-coupling} in the small-coupling limit, in the far-field limit, $r \gg M$ (we have just reversed the order in which the limits are taken). We obtain, up to corrections of relative order $\zeta^2$:
\barr
\vartheta_1(r \geq R \gg M, \zeta \ll 1) &\approx& \frac{J D_1}{\sqrt{2 \pi} r^2} - 4 \frac{\ell_{\rm cs}^2 M J}{r^5}, \label{eq:theta1-full-expansion}\\
\omega(r \geq R \gg M, \zeta \ll 1) &\approx&  \frac{2 J}{r^3} - 8 \sqrt{2 \pi} \frac{J \ell_{\rm cs}^2 M D_1}{r^6} \nonumber\\
 &&+ \frac{128 \pi}{3} \frac{J \ell_{\rm cs}^4 M^2}{r^9}. \label{eq:omega-full-expansion}
\earr
Comparing to Eqs.~(\ref{eq:theta-expansion}) and (\ref{eq:domega far}), we see that the constants $C_1$ and $D_1$ are related in the regime $M\ll R, \zeta \ll1$:
\beq
D_1 = \frac{5 \sqrt{2 \pi}}{8} \frac{\ell_{\rm cs}^2}{M^2}(1 - C_1).
\eeq
Moreover, inspection of the second terms in Eqs.~(\ref{eq:theta-expansion}) and (\ref{eq:theta1-full-expansion}) show that one must have
\beq
C_1 = 1 + \mathcal{O}(M/R) + \mathcal{O}(\zeta^2).
\eeq

\subsection{Analytic solution in the nonrelativistic limit for a constant density object} \label{sec:analytic-constant-density}

We have already obtained the general solution outside a nonrelativistic star in the previous section. We now need a solution inside the star to obtain the integration constants $D_1$ and $J$. 

\begin{figure}
\includegraphics[width = 85 mm]{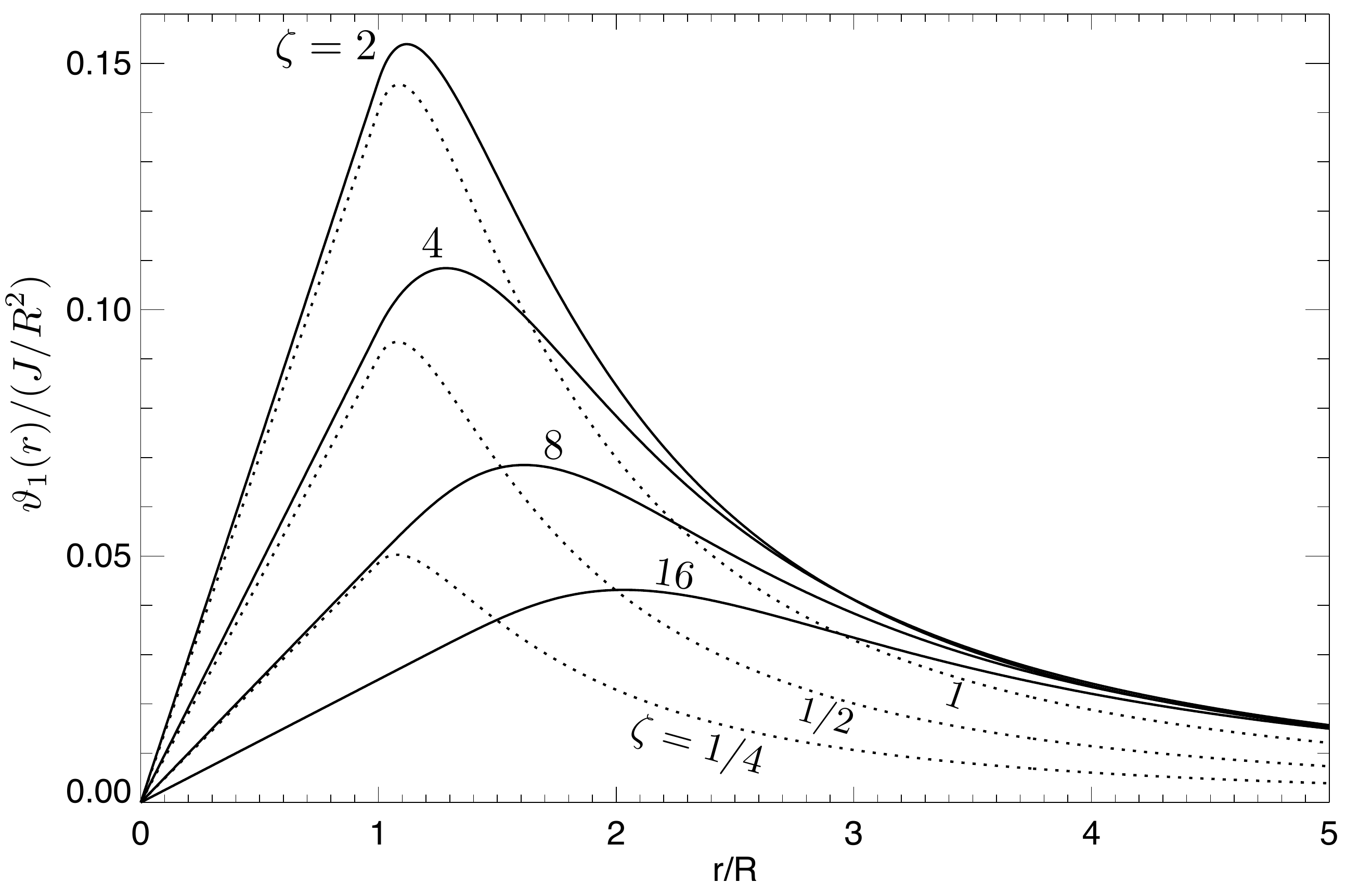}
\caption{Chern-Simons scalar field $\vartheta_1$ as a function of radius, for a constant-density nonrelativistic object. The curves are parametrized by the dimensionless coupling constant $\zeta$ defined in Eq.~(\ref{eq:zeta}). For the ease of visualization, we have used dotted lines for $\zeta \leq 1$ (the linear or quasi-linear regime, where $\vartheta_1$ increases uniformly with $\zeta$), and solid lines for $\zeta > 1$ where non-linearity results into a damping of the scalar field in the near-zone and a plateauing of its asymptotic behavior at large radii.} \label{fig:theta-earth} 
\end{figure}

Let us first consider Eq.~(\ref{eq:ODE-theta1}) for $\vartheta_1$. For a constant density object, the right-hand-side vanishes, and in the non-relativistic limit, we find 
\beq
\vartheta_1(r \leq R) = C_3 \frac r R + C_4\frac{R^2}{r^2}.
\eeq
Requiring $\vartheta_1$ to be finite at the center of the star, we set $C_4 = 0$. We therefore have $R \vartheta_1'(R-\epsilon) = \vartheta_1(R-\epsilon)$. The continuity and smoothness of $\vartheta_1$ at the boundary therefore imply that this relation is also satisfied at $R+ \epsilon$. Imposing this condition with $\vartheta_1$ given in Eq.~(\ref{eq:theta1-out-nonrel}) (where we recall that $D_2 = 1$), we obtain, up to corrections of order $M/R$, 
\barr
C_3 &=& \frac{J}{\sqrt{2 \pi} \zeta R^2} \left(1 - \frac{1}{\cosh \zeta}\right),\\
D_1 &=& \tanh \zeta.
\earr
From the value of $D_1$ and Eq.~(\ref{eq:theta1-full-expansion}) we infer the asymptotic behavior of $\vartheta_1$ at large radii ($r \gg R \zeta^{1/3}$):
\beq
\frac{r^2}{R^2} \vartheta_1 \underset{r \rightarrow \infty}{\rightarrow} \frac{J}{\sqrt{2 \pi} R^2} \tanh \zeta. \label{eq:theta-far}
\eeq
We show the radial dependence of the scalar field $\vartheta_1(r)$ for a nonrelativistic constant density object in Fig.~\ref{fig:theta-earth}. We see that in the linear regime ($\zeta \lesssim 1$), the scalar field increases uniformly with $\zeta$ --- for $\zeta \ll 1$, $\vartheta_1 \propto \zeta^2$. Increasing $\zeta$ further eventually leads to a damping of the scalar field near the surface of the star. In the far zone, we see from Eq.~(\ref{eq:theta-far}) that the asymptotic value of $r^2 \vartheta_1$ plateaus to a constant value.

\begin{figure}
\includegraphics[width = 85 mm]{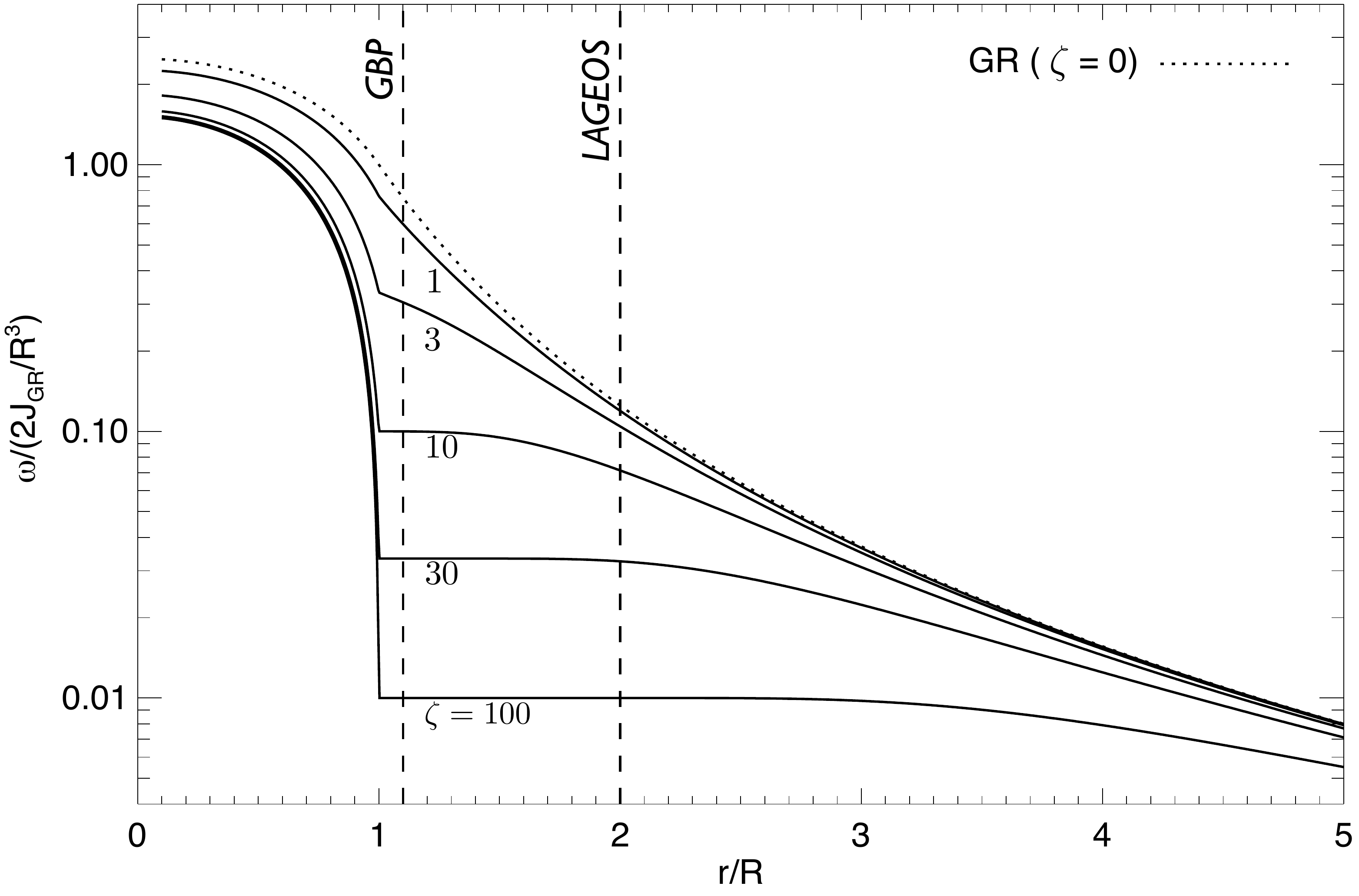}
\caption{Function $\omega(r)$ in CS gravity around a constant-density nonrelativistic object, for various values of the dimensionless coupling strength $\zeta$ defined in Eq.~(\ref{eq:zeta}), marked as labels. The dotted curve corresponds to standard GR ($\zeta = 0$). Because of the sharp boundary of a constant-density object, this function is continuous but not smooth, and the jump in its derivative at the surface is given by Eq.~(\ref{eq:jump}). All curves asymptote to $\omega \approx 2J/r^3$ for $r \gg R \zeta^{1/3}$ but differ significantly close to the surface. The vertical dashed lines show the locations probed by the LAGEOS \cite{Lageos} and GPB \cite{GPB} satellites, which orbit the Earth at an altitude of $6000$ km and $640$ km, respectively, and are used to set constraints to the CS lengthscale in Sec.~\ref{sec:constraints}.} \label{fig:earth} 
\end{figure}

Let us now consider Eq.~(\ref{eq:ODE-omega1}) for $\omega$. Again, for a constant density object, the right-hand-side vanishes. Using $\rho = 3M/(4 \pi R^3)$, the equation satisfied by $\omega$ becomes, to first order in $M/R$:
\beq
\left(1 - \frac{2 M}{R^3}r^2\right) \omega'' + \frac{4}{r}\left(1 - \frac{11 M}{4 R^3}r^2 \right) \omega' + \frac{12 M}{R^3} (\Omega - \omega) = 0, 
\eeq
which has the general solution, up to corrections of order $(M/R)^2$:
\beq
\omega = \Omega + D_3\left(1 + \frac{6 M}{5 R^3} r^2\right) + \frac{D_4}{r^4}\left( 1 - \frac{4 M}{R^3} r^2\right).
\eeq
In order for $\omega$ to be finite at the origin, we impose $D_4 = 0$. Requiring $\omega$ to be continuous at the boundary and its derivative to satisfy the jump condition Eq.~(\ref{eq:jump}), we find $D_3 = - \frac52 J/(M R^2)$ and
\beq
J = \frac{2}{5} MR^2 \Omega\left[1 - \frac{2 M}{R} + \frac{4 M}{5 R}\left(1- \frac{\tanh \zeta}{\zeta} \right)  + \mathcal{O}(M/R)^2\right].
\eeq
If we write $J = J_{\rm GR} + \Delta J_{\rm CS}$ (with the well-known result $J_{\rm GR} = \frac25 MR^2 \Omega$ in the non-relativistic limit), we therefore obtain, to lowest order in $M/R$,
\beq
\frac{\Delta J_{\rm CS}}{J_{\rm GR}} = \frac{4 M}{5 R}\left( 1 - \frac{\tanh \zeta}{\zeta}\right).
\eeq
In the small coupling limit ($\zeta \ll 1$), we find
\beq
\frac{\Delta J_{\rm CS}}{J_{\rm GR}} \underset{\zeta \ll 1}{\approx} \frac{4M}{5R} \frac{\zeta^2}{3} = \frac{512 \pi}{15} \frac{\ell_{\rm cs}^4 M^3}{R^7}. \label{eq:DJ_J-const-rho-small}
\eeq
Interestingly, for large values of the coupling constant, the Chern-Simons correction saturates
\beq
\frac{\Delta J_{\rm CS}}{J_{\rm GR}} \underset{\zeta \gg 1}{\approx} \frac{4M}{5R}. 
\eeq
We show the function $\omega(r)$ for a constant-density, nonrelativistic object in Fig.~\ref{fig:earth}. The effect of CS gravity is to (potentially strongly) decrease the gravitomagnetic potential near the surface, and enhance its asymptotic value in the far field $r \gg R \zeta^{1/3}$ by a relative amount $\Delta J_{\rm CS}/J_{\rm GR} < 4M/(5R) \ll 1$.

In the next section we will see that the qualitative features of the analytic solution are recovered in the full numerical solution.




\section{Numerical solution for realistic neutron stars} \label{sec:numerical}

In the previous section we have given analytic solutions for the coupled CS scalar field-metric system in the nonrelativistic limit, for constant density objects. In the present section we provide full numerical solutions for neutron stars described by realistic equations of state.

\subsection{Equations of state}

\begin{figure}
\includegraphics[width = 85 mm]{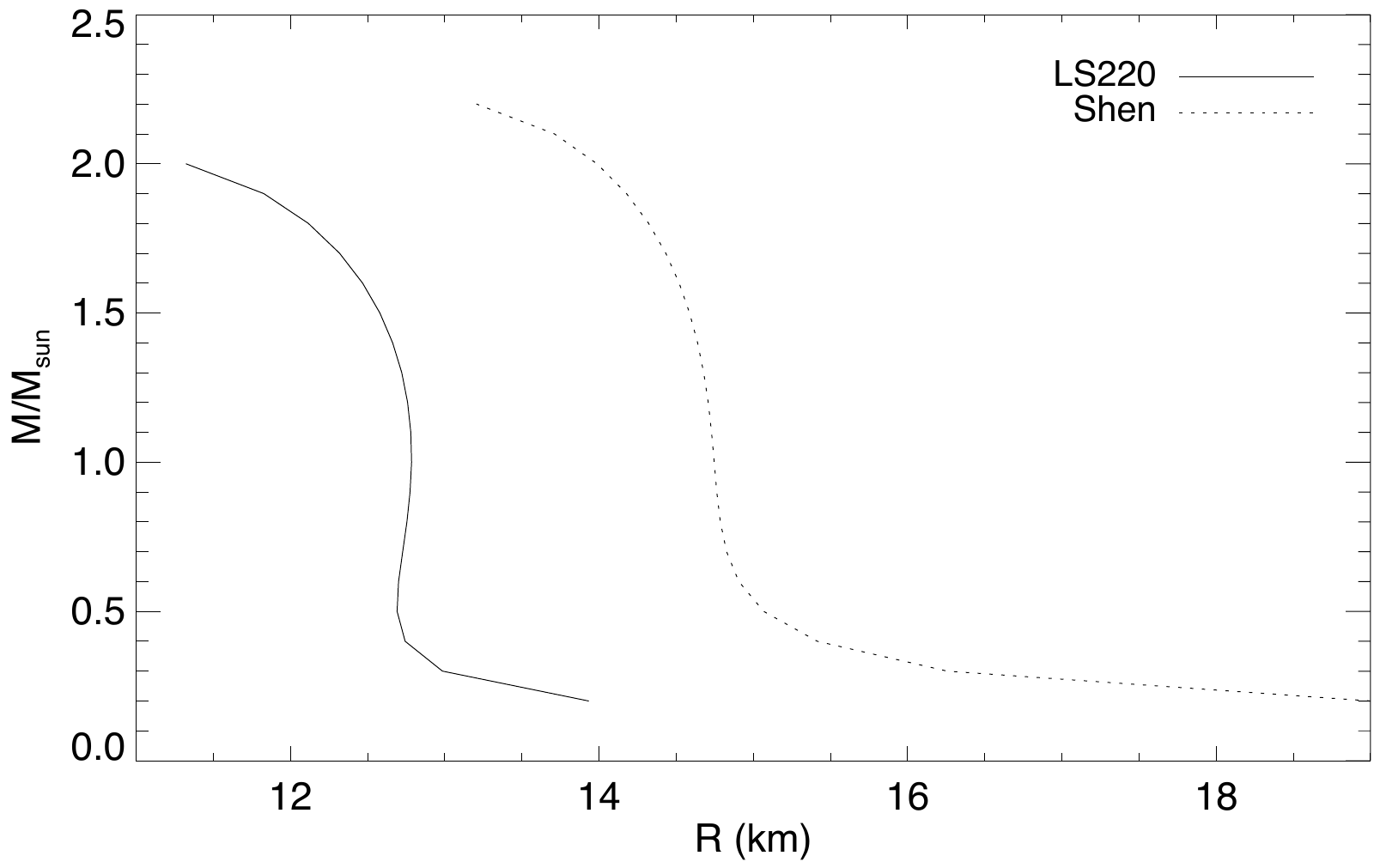}
\caption{Neutron star mass-radius relation for the two equations of state used in this work: Lattimer-Swesty (LS220) \cite{LS_EOS} and Shen \emph{et al} \cite{Shen_1, Shen_2}.} \label{fig:M-R} 
\end{figure}

Matter at nuclear densities has complex properties, and the EOS of neutron stars is not very well known. There exist two EOSs that are widely used in astrophysical simulations (see for example Ref.~\cite{Evan} for a discussion): the Lattimer-Swesty (LS) EOS \cite{LS_EOS} and the Shen \emph{et al} EOS \cite{Shen_1, Shen_2}. We will use the LS EOS with nuclear incompressibility $K_0 = 220$ MeV (hereafter, LS220) and the Shen \emph{et al} EOS (hereafter Shen). We use the EOS routines of O'Connor and Ott\footnote{Available at http://stellarcollapse.org/microphysics; we thank Evan O'Connor for providing tabulated solutions to the relativistic equations of stellar structure.} \cite{Evan} to solve the relativistic stellar structure equations. We show the resulting mass-radius relations in Fig.~\ref{fig:M-R}. 

\begin{figure}
\includegraphics[width = 85 mm]{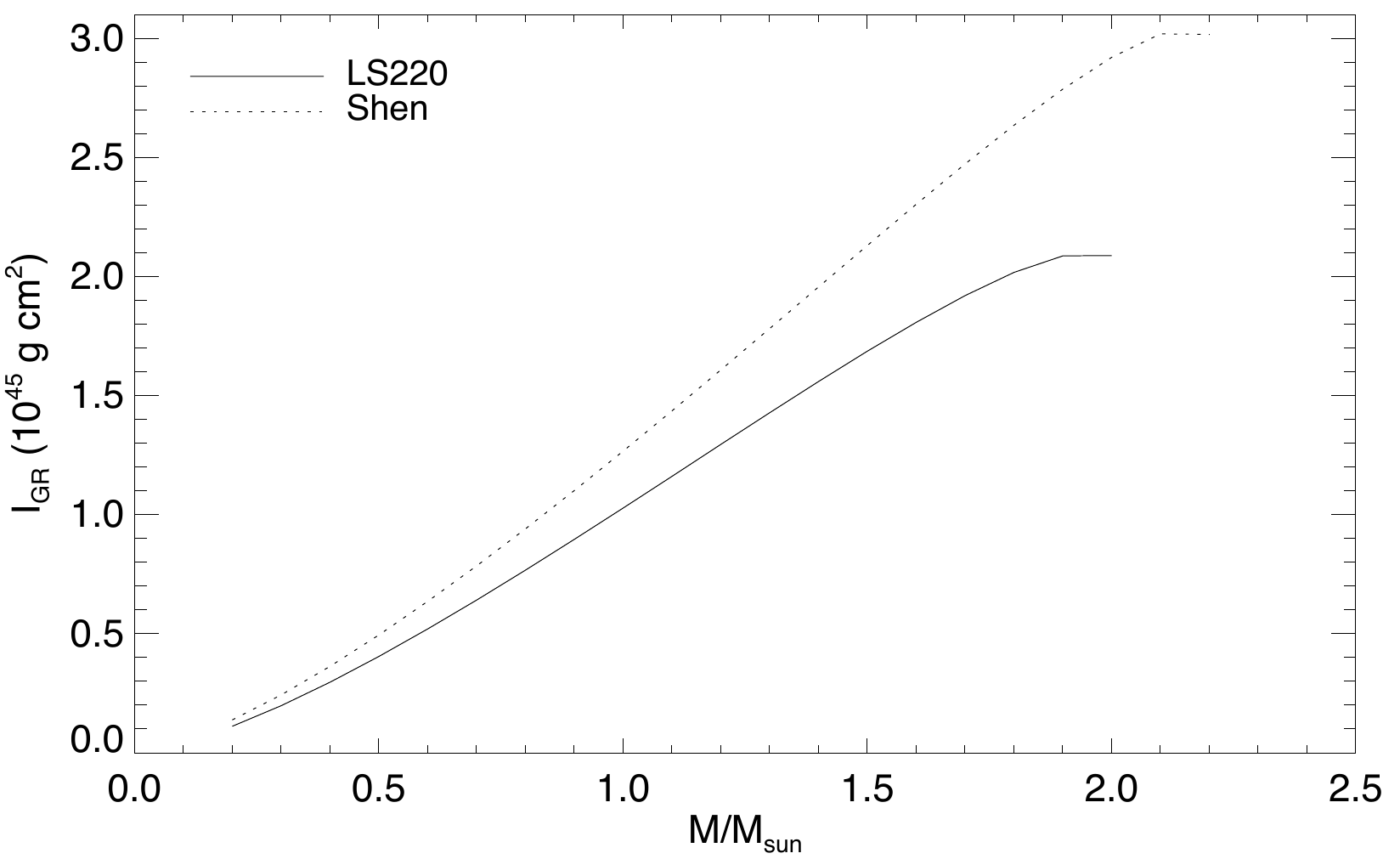}
\caption{Moment of inertia as a function of neutron star mass in standard GR, for the two equations of state used in this work: Lattimer-Swesty (LS220) \cite{LS_EOS} and Shen \emph{et al} \cite{Shen_1, Shen_2}.} \label{fig:inertia} 
\end{figure}

We also show the moment of inertia in standard GR as a function of neutron star mass and equation of state in Fig.~\ref{fig:inertia}. To evaluate the moment of inertia, we have integrated Eq.~(\ref{eq:omega_Hartle}) in the absence of CS coupling ($\ell_{\rm cs}^2 = 0$) with a second-order implicit Euler method.

\subsection{Numerical solution of the coupled CS equations}

\subsubsection{Boundary conditions}

Inspection of the system given by Eqs.~(\ref{eq:ODE-omega}) and (\ref{eq:ODE-theta1}) near the origin shows that there is one well behaved solution for $\omega$ with $\omega(r) \underset{r \rightarrow 0}{\sim} c_1 + \mathcal{O}(r^2)$ and a divergent (hence unphysical) solution with $\omega(r) \underset{r \rightarrow 0}{\sim} c_2/r^3$. Similarly, there is one well-behaved solution for $\vartheta_1$ with $\vartheta_1(r) \underset{r \rightarrow 0}{\sim} c_3 r$ and a divergent, unphysical solution with $\vartheta_1 \underset{r \rightarrow 0}{\sim} c_4/r^2$. The physically allowed boundary conditions at the origin are therefore
\barr
\omega(0) &=& c_1, \ \ \omega'(0) = 0, \\
\vartheta_1(0) &=& 0, \ \ \vartheta_1'(0) = c_3,
\earr
where $c_1$ and $c_3$ are integration constants to be determined.

The mathematically allowed asymptotic behaviors at infinity are similar, but the physically relevant solutions are reversed, i.e. we have
\barr
\lim_{r \rightarrow \infty} r^3\omega(r) = 2 J, \ \ \lim_{r \rightarrow \infty} \frac{\rmd}{\rmd r} [r^3\omega(r)] = 0,\\
\lim_{r \rightarrow \infty} r^2\vartheta_1(r) = d_4, \ \ \lim_{r \rightarrow \infty} \frac{\rmd}{\rmd r} [r^2\vartheta_1(r)] = 0,
\earr
where $d_4$ and $J$ are integration constants, and $J = J_{\rm GR} + \Delta J_{\rm CS}$ can physically be interpreted as the total perceived angular momentum of the system.

In addition to these boundary conditions at $r=0$ and $+\infty$, the functions $\omega, \omega', \vartheta_1$ and $\vartheta_1'$ must all be continuous at the boundary of the star (in principle there is a jump in the derivative of $\omega$, see Eq.~(\ref{eq:jump}), but for realistic neutron stars the surface density is 7 orders of magnitude than the mean density, so $\omega'$ is continuous up to small corrections).

\subsubsection{Shooting method}

We integrate the coupled system given by Eqs.~(\ref{eq:ODE-omega}) and (\ref{eq:ODE-theta1}) with a second-order implicit Euler method from $r = 0$ to $R$, and from $r = +\infty$ down to $r=R$ (specifically, we consider the functions $\tilde{\omega}(u) \equiv \frac{1}{u^2} \omega(1/u)$ and $\tilde{\vartheta_1}(u) \equiv \frac{1}{u} \vartheta_1(1/u)$ from $u = 0$ to $1/R$; these functions vanish at $u = 0$ and their derivatives at $u= 0$ are proportional to the constants $J$ and $d_4$). The linearity of the system allows to find the appropriate constants of integration at $r = 0$ and $+\infty$ with a shooting method by requiring continuity and smoothness at $r = R$.

\subsection{Results}

In this section we illustrate our numerical results in several figures, and compare them to our analytic solution of Sec.~\ref{sec:analytic-constant-density}.

\subsubsection{Scalar field}

\begin{figure}
\includegraphics[width = 85 mm]{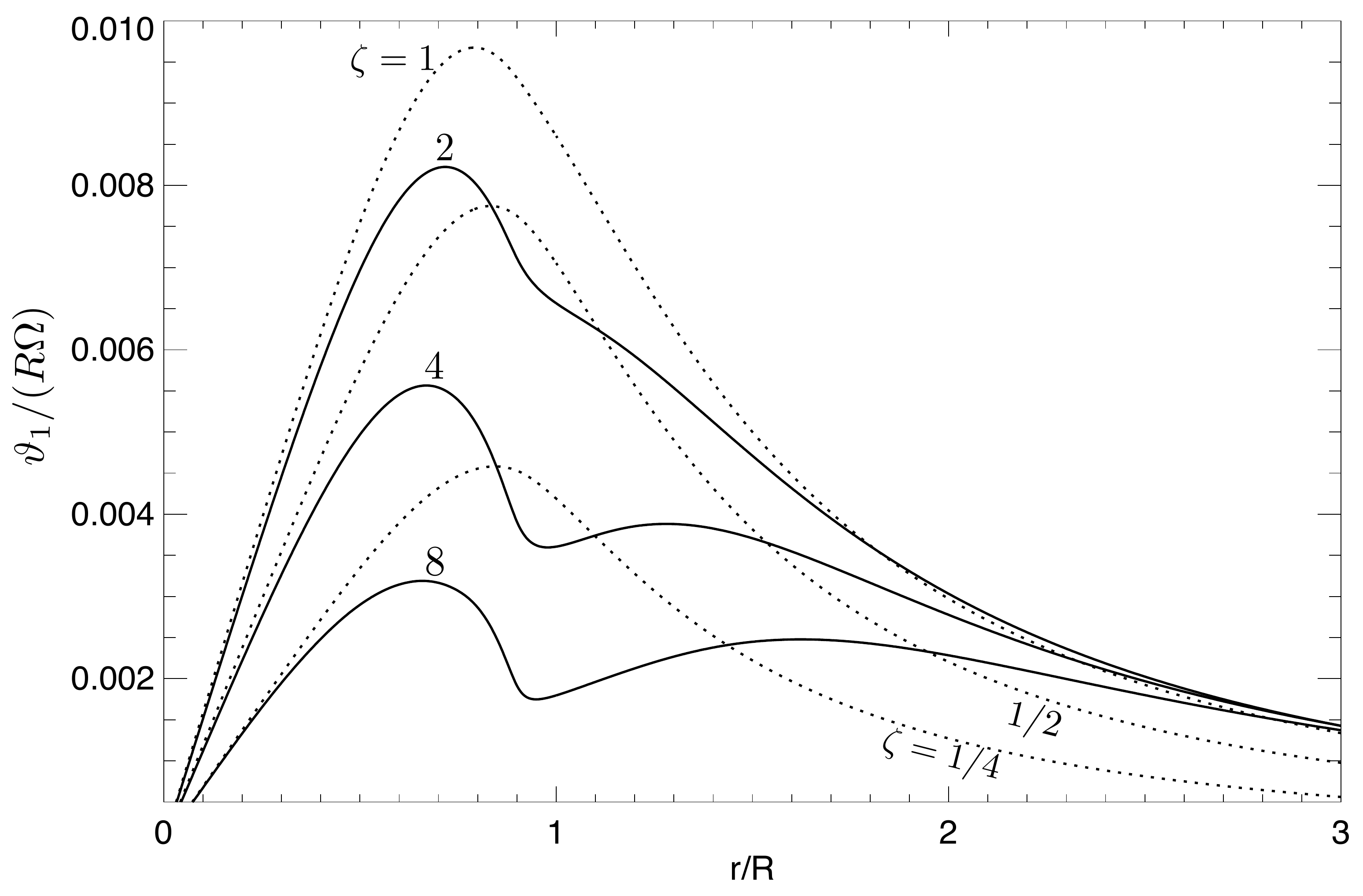}
\caption{Radial dependence of the Chern-Simons scalar field $\vartheta_1(r)$ for a 1 $M_{\odot}$ neutron star described by the Shen et al EOS (the LS220 EOS shows a very similar behavior and we have not plotted it here for more clarity). The curves are labelled by the dimensionless coupling strength $\zeta$ defined in Eq.~(\ref{eq:zeta}). For $\zeta \lesssim 1$, the scalar field increases uniformly over the whole range $r > 0$. For $\zeta \gtrsim 1$, the scalar field gets suppressed in the vicinity of the star; in the far field, $\underset{r \rightarrow\infty}{\lim}r^2 \vartheta_1(r)$ asymptotes to a constant value for large values of $\zeta$. We have shown the two regimes $\zeta \leq 1$ and $\zeta > 1$ with different line styles for more visual clarity.
} \label{fig:theta1} 
\end{figure}

First, in Fig.~\ref{fig:theta1}, we show the scalar field $\vartheta_1(r)$ as a function of radius, for several values of the dimensionless CS coupling parameter $\zeta$. A qualitative difference of this work with that of Ref.~\cite{Yunes_et_al_10} is that we have properly enforced the continuity \emph{and} the smoothness of the CS scalar field at the stellar boundary (see Fig.~4 of Ref.~\cite{Yunes_et_al_10} for a comparison). In the linear regime $\zeta \lesssim 1$, $\vartheta_1$ increases uniformly with $\zeta$. For $\zeta \gtrsim 1$, more complex nonlinear behaviors appear (the system solved being equivalent to a fourth-order ODE, such behaviors can be expected), but the overall effect is to suppress $\vartheta$ near the stellar surface. The far-field behavior of $\vartheta$ plateaus to an asymptotic limit, as we found in our analytic approximation.

\subsubsection{Gravitomagnetic sector and moment of inertia}

\begin{figure}
\includegraphics[width = 85 mm]{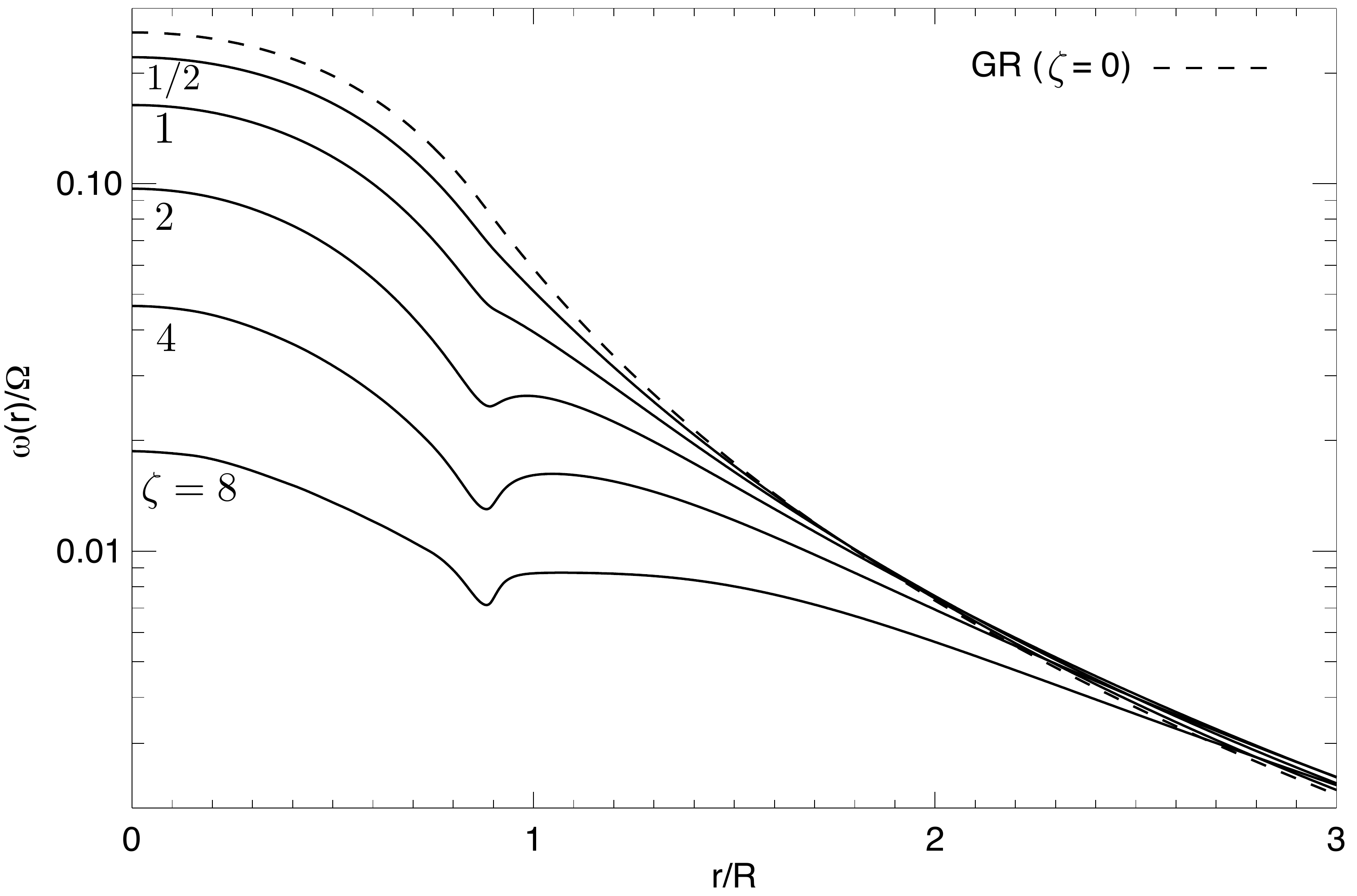}
\caption{Radial dependence of the normalized metric coefficient $\omega(r)/\Omega$, for a 1 $M_{\odot}$ neutron star described by the Shen et al EOS (the LS220 EOS shows a very similar behavior and the corresponding result was not plotted here). The curves are labelled by the dimensionless coupling strength $\zeta$ defined in Eq.~(\ref{eq:zeta}). For $\zeta \gtrsim 1$ frame dragging becomes strongly suppressed in the vicinity of the star. At large radii $\omega \approx 2J/r^3$, where $J = J_{\rm GR} + \Delta J_{\rm CS}$ is enhanced with respect to the value in GR.} \label{fig:Domega} 
\end{figure}

In Fig.~\ref{fig:Domega}, we show the gravitomagnetic sector of the metric through the function $\omega(r)$. An essential difference of our solution with previous work is that we have $\Delta \omega_{\rm CS} \approx 2 \Delta J_{\rm CS}/r^3$ at large radii, instead of $\Delta \omega_{\rm CS} \propto 1/r^6$.

\begin{figure}
\includegraphics[width = 85 mm]{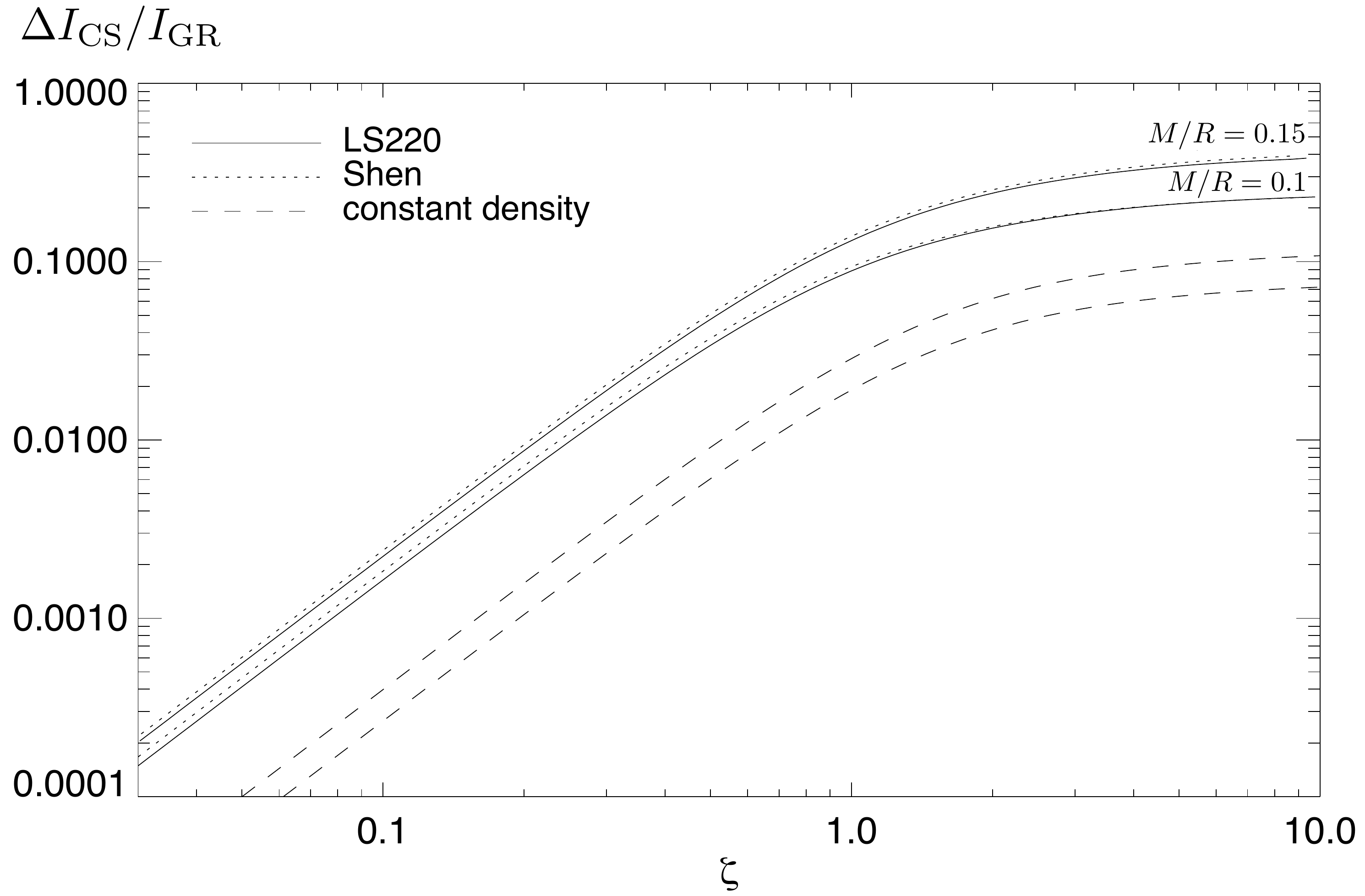}
\caption{Change in the moment of inertia $\Delta I_{\rm CS}/I_{\rm GR}$ induced by the CS modification as a function of the CS coupling strength $\zeta$, for the two EOSs considered. We plot $\Delta I_{\rm CS}/I_{\rm GR}$ for the values $M/R = 0.1$ and 0.15. The dashed lines show the analytic result derived in Sec.~\ref{sec:analytic-constant-density} for a constant density nonrelativistic star (also for $M/R = 0.1$ and 0.15 from bottom to top). } \label{fig:DI-full} 
\end{figure}

In Fig.~\ref{fig:DI-full} we show the relative change in moment of inertia induced by the CS modification, $\Delta I_{\rm CS}/I_{\rm GR}$, as a function of the dimensionless coupling strength $\zeta$. We recall that the moment of inertia is defined as a $I \equiv J/\Omega$, where $J = \frac12 \underset{r \rightarrow \infty}{\lim}[r^3 \omega(r)]$. We also plot the result of our analytic approximation for a nonrelativistic constant density star. Although the overall normalization is off by nearly an order of magnitude, we see that the trends predicted by our simple approximation are indeed recovered in the full numerical result. In the limit of small CS coupling, we have $\Delta I_{\rm CS}/I_{\rm GR} \propto \zeta^2 \propto \ell_{\rm cs}^4$. For large values of the dimensionless coupling parameter, the CS correction to the moment of inertia asymptotes to a constant value.

\begin{figure}
\includegraphics[width = 85 mm]{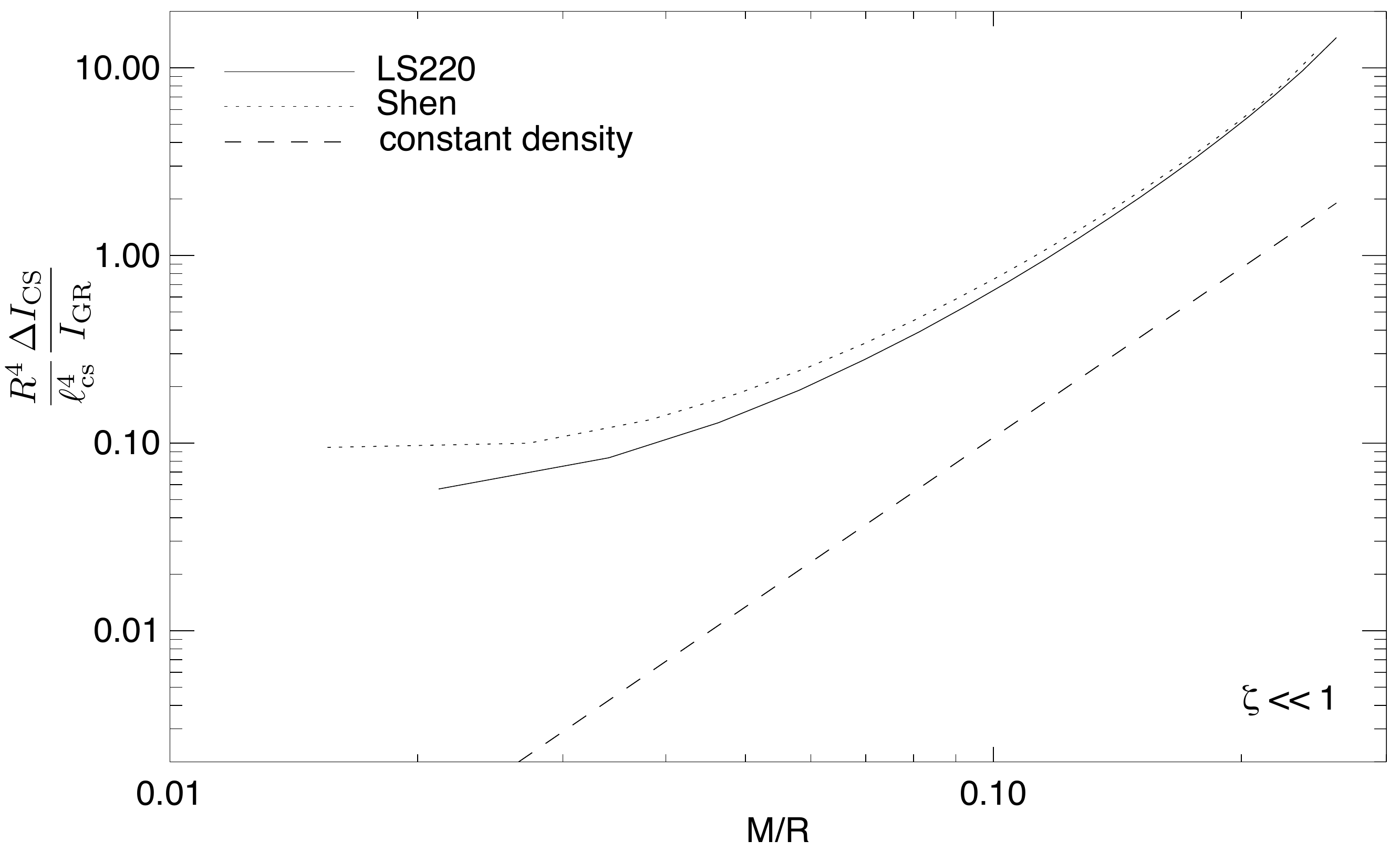}
\caption{Relative change in the moment of inertia $\Delta I_{\rm CS}$ induced by the CS modification in the small coupling limit, as a function of neutron star compactness $M/R$, for the two EOSs considered. We show the dimensionless quantity $\lim_{\zeta \rightarrow 0} R^4/\ell_{\rm cs}^4 \Delta I_{\rm CS}/I_{\rm GR}$. The dashed line shows the analytic result derived in Sec.~\ref{sec:analytic-constant-density} for a constant density nonrelativistic star. The amplitude is off by approximately an order of magnitude but the overall behavior is relatively well reproduced.} \label{fig:DI-small} 
\end{figure}

We show the dependence of $\Delta I_{\rm CS}/I_{\rm GR}$ on neutron star compactness in the small coupling limit in Fig.~\ref{fig:DI-small}. Again, our very simple analytic result is underestimating the exact numerical result, but the trend $\Delta I_{\rm CS}/I_{\rm GR} \propto (M/R)^3$ is roughly respected for realistic neutron star masses. 

\begin{figure}
\includegraphics[width = 85 mm]{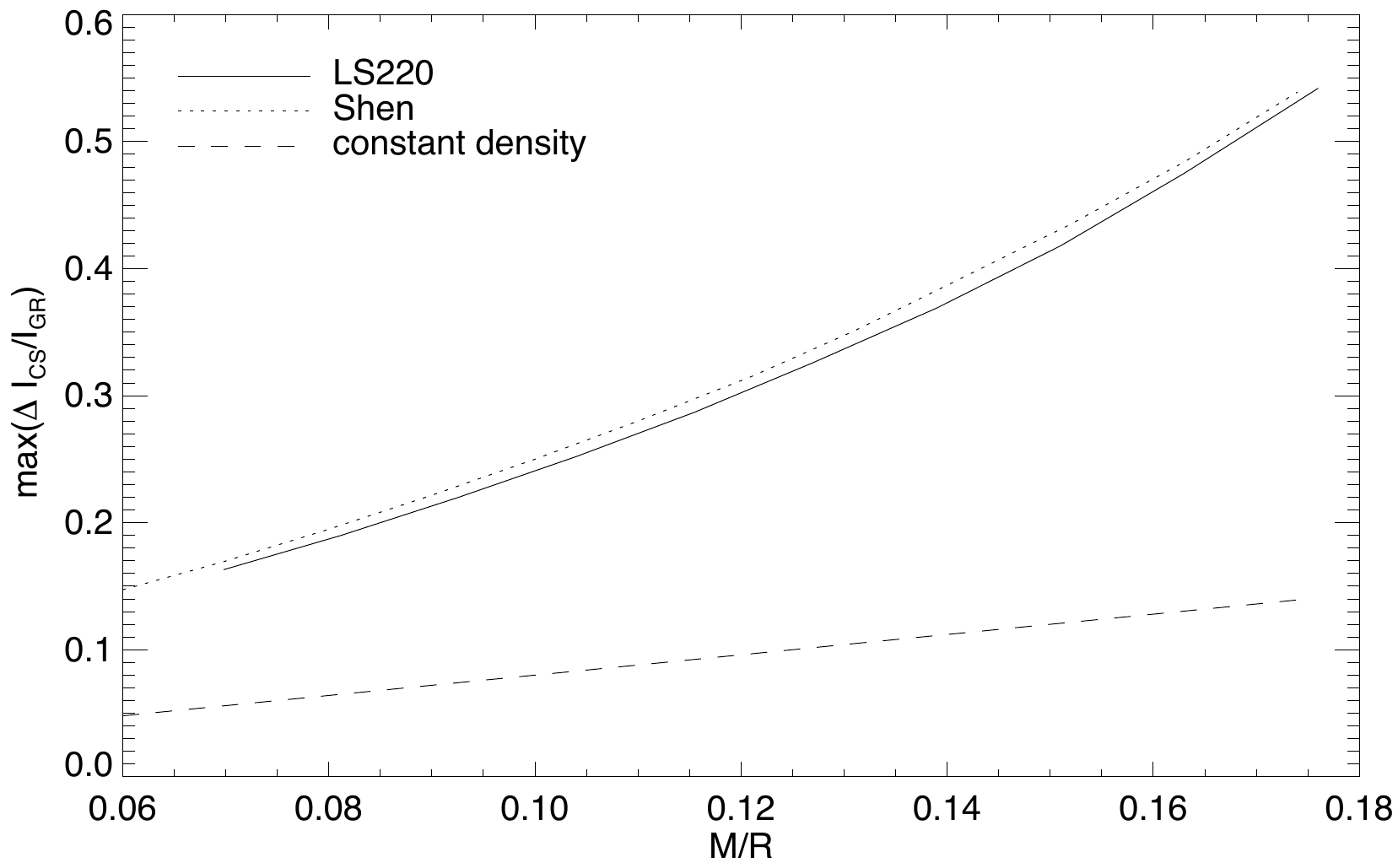}
\caption{Asymptotic value of the CS change in moment of inertia, $\max(\Delta I_{\rm CS}/I_{\rm GR})$, as a function of neutron star compactness. Also shown is our analytic approximation for nonrelativistic constant density objects.} \label{fig:DI-large} 
\end{figure}

Finally, in Fig.~\ref{fig:DI-large} we show the extrapolated asymptotic value of $\Delta I_{\rm CS}/I_{\rm GR}$ in the limit of large coupling parameter, as a function of neutron star compactness. It depends approximately linearly on neutron star compactness, as we found in our analytic solution. To obtain this value we have computed $\Delta I_{\rm CS}/I_{\rm GR}(\zeta)$ for several values of the coupling strength up to $\zeta \approx 10$ and fitted $\Delta I_{\rm CS}/I_{\rm GR} = \Delta I_{\rm CS}/I_{\rm GR}|_{\infty}\left(1 - \alpha/\zeta\right)$ --- we could not go much beyond $\zeta \approx 10$ as the homogeneous solutions contain terms of order $\exp(\zeta)$ that quickly become very large and lead to large numerical errors, see equations of Sec.~\ref{sec:analytic-zeta}.

Our numerical results are therefore in very good qualitative agreement with our analytic approximation for constant-density nonrelativistic objects. Given that a neutron star density profile is far from constant and that neutron stars are compact, it is not surprising that the quantitative agreement is not perfect. We can understand why the analytic approximation systematically underestimates the correct result. A realistic neutron star is much more centrally concentrated than a constant density object. The ``effective'' radius of the star (defined, for example, as the radius containing 75\% of the mass), is always significantly smaller than the actual stellar radius. A neutron star is therefore more compact in its central regions than if it had a constant density. For example, the mean density of a 1 $M_{\odot}$ neutron star inside the sphere containing 75\% of its mass is about 70\% larger than its overall mean density. 

\section{Slowly-rotating black hole in Chern-Simons gravity} \label{sec:black hole}

In this section we compute the solution for the scalar field and the metric of a slowly-rotating black hole in CS gravity, for an arbitrary coupling constant. 

The equation satisfied by the CS scalar field for a slowly-rotating black hole is the same as that outside a slowly-rotating star, Eq.~(\ref{eq:ODE-theta-out}). Using $x \equiv r/(2M)$, and defining the dimensionless coupling parameter in analogy with Eq.~(\ref{eq:zeta}) with the substitution $R = 2M$
\beq
\zeta_{\rm BH}^2 \equiv 2\pi \frac{\ell^4_{\rm cs}}{M^4},
\eeq
the equation for the scalar field becomes
\barr
\left(1 - \frac1x\right)\ddot{\vartheta}_1 + \frac2x\left(1 - \frac1{2x}\right) \dot{\vartheta}_1 - \left(\frac2{x^2} + \frac{9 \zeta_{\rm BH}^2}{x^8} \right) \vartheta_1 \nonumber\\
= - \frac94 \frac{\ell^2_{\rm cs} J}{M^4 x^7},~~~ \label{eq:ODE-theta-bh}
\earr
where dots denote differentiation with respect to $x$.

There are two qualitative differences between the black hole case and the stellar case. 

Firstly, in the case of a slowly-rotating black hole, the angular momentum $J$ is just a given parameter (as well as the black hole mass), independently of the gravity theory chosen. Of course, $J$ is the angular momentum of the object that collapsed into a black hole, in which case its value does depend on the gravity theory chosen; once the black hole is formed, however, there is no way to disentangle $J_{\rm GR}$ from $J_{\rm CS}$. 

Secondly, the boundary conditions are different than in the case of a star. For a black hole, the physical solution for $\vartheta$ must be continuous and smooth at the horizon $r = 2M$. Since there is a homogeneous solution that behaves as $\vartheta \sim \ln(1 - 2M/r)$ near the horizon [see Eq.~(\ref{eq:outside-sol})], we must chose the boundary conditions in a way to avoid such a logarithmic divergence.

To numerically solve Eq.~(\ref{eq:ODE-theta-bh}), we start near the singular point $x =1$ with initial conditions 
\barr
\vartheta_1(1+ \epsilon) &=& C,\nonumber\\
\dot{\vartheta}_1(1 + \epsilon) &=& - \frac94 \frac{ \ell_{\rm cs}^2J}{M^5} + \left(2 + 9 \zeta_{\rm BH}^2\right)C,
\earr
where $C$ is a constant to be determined and the second equation follows from Eq.~(\ref{eq:ODE-theta-bh}) evaluated at $x = 1$. We then integrate Eq.~(\ref{eq:ODE-theta-bh}) outward to some radius $x_0 >1$. We also integrate Eq.~(\ref{eq:ODE-theta-bh}) from $+\infty$ to $x_0$, with initial conditions $\underset{x \rightarrow \infty}{\lim}[x^2 \vartheta_1(x)] = D$, where $D$ is a constant. We then adjust $C$ and $D$ such that $\vartheta_1$ is continuous and smooth at the junction radius $x_0$. Finally, $\omega$ is obtained from Eq.~(\ref{eq:omega'-out}).

\begin{figure}
\includegraphics[width = 85 mm]{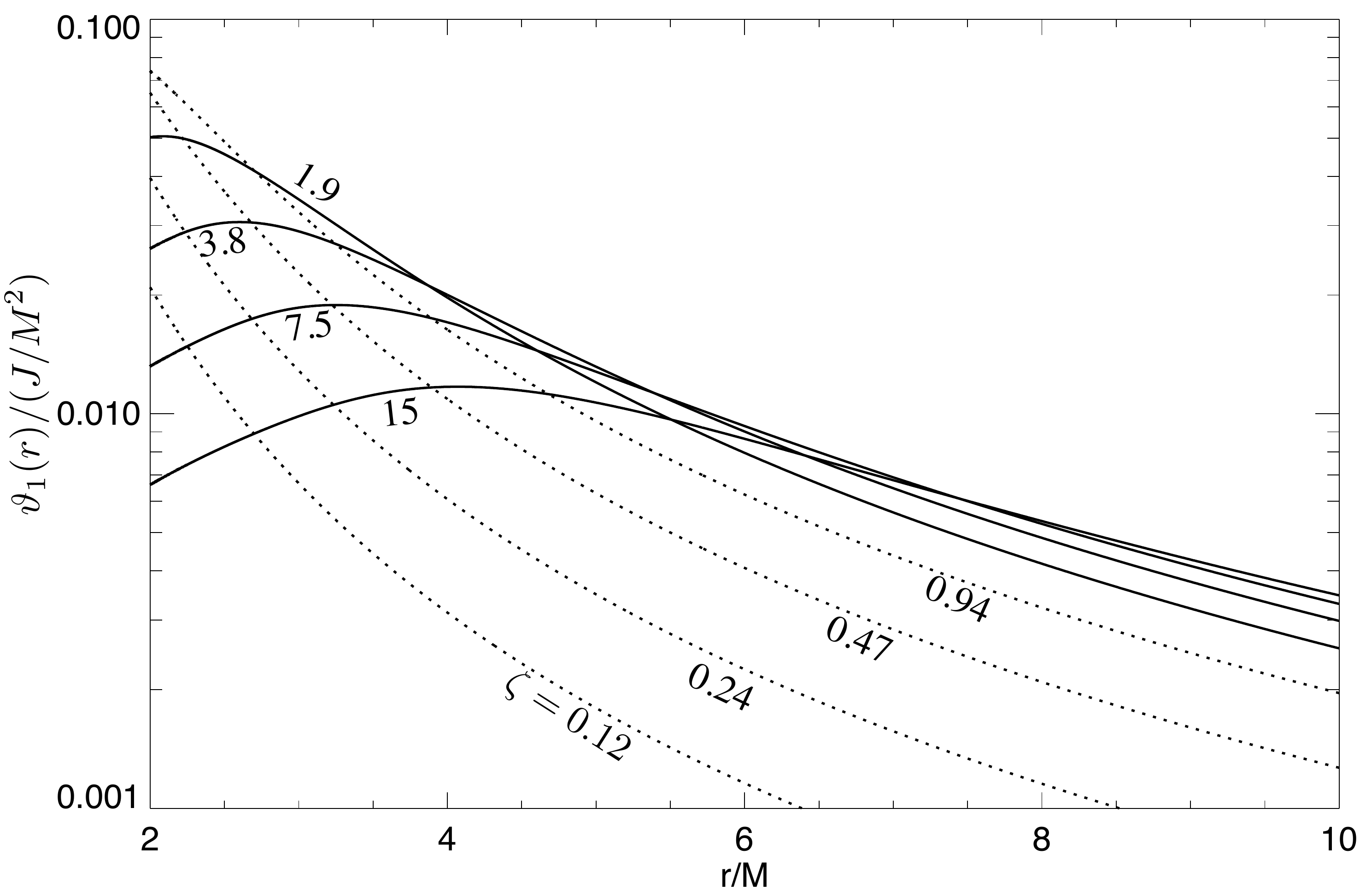}
\caption{Chern-Simons scalar field outside the horizon of a slowly rotating black hole, for various values of the dimensionless coupling parameter $\zeta \equiv \sqrt{2\pi} \ell_{\rm cs}^2/M^2$. For better clarity, we have shown the small-coupling regime $\zeta \leq 1$ with dotted lines and the nonlinear regime $\zeta > 1$ with solid lines.} \label{fig:theta-app} 
\end{figure}

\begin{figure}
\includegraphics[width = 85 mm]{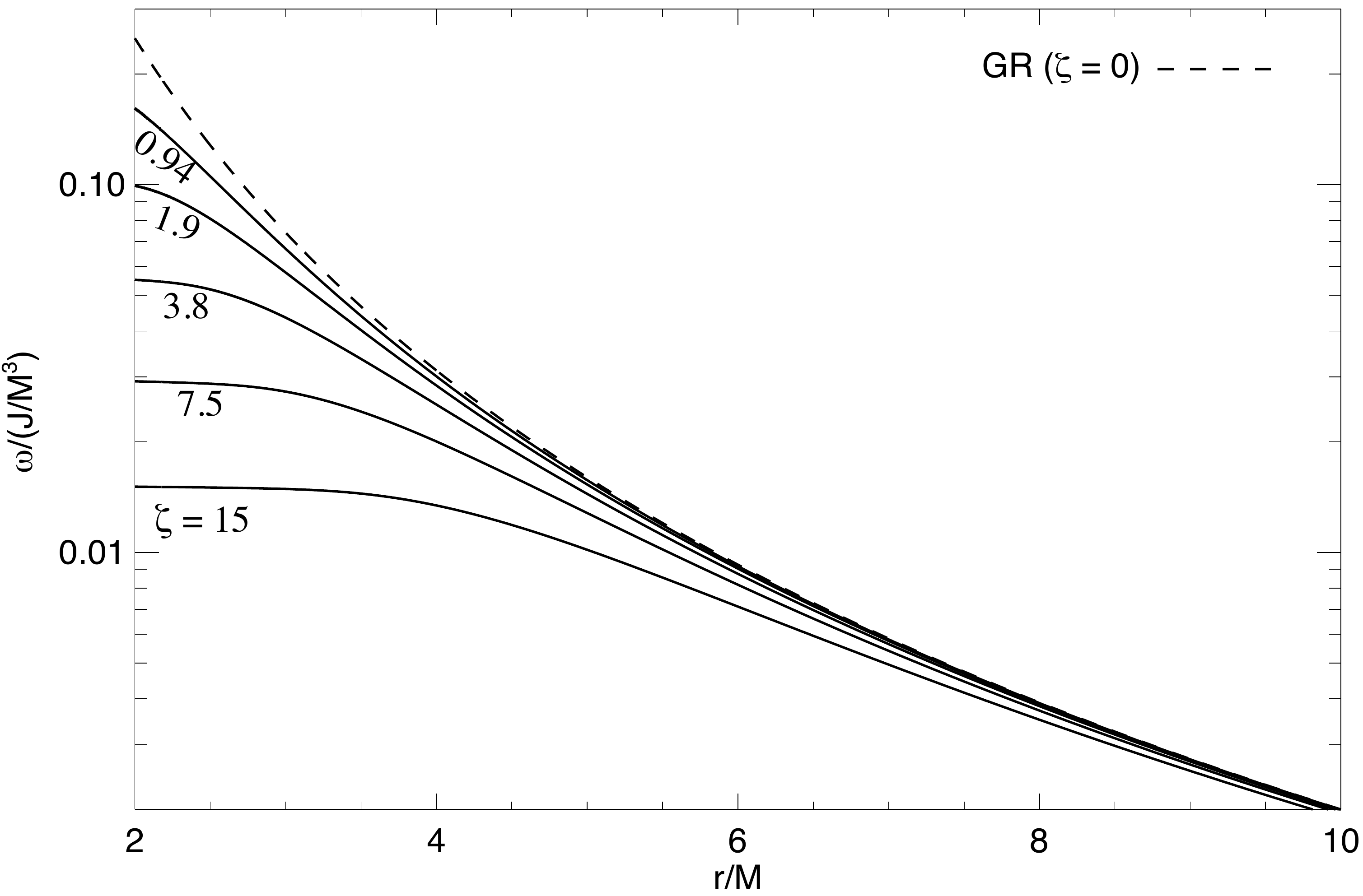}
\caption{Metric coefficient $\omega(r)$ outside the horizon a slowly rotating black hole, for various values of the dimensionless CS coupling parameter $\zeta \equiv \sqrt{2\pi} \ell_{\rm cs}^2/M^2$.} \label{fig:omega-app} 
\end{figure}

The solution scales linearly with the dimensionless parameter $J/M^2$ and otherwise depends on the single parameter $\zeta_{\rm BH}$. We checked that in the small-coupling limit $\zeta_{\rm BH} \ll 1$ our numerical solution agrees with the analytic result of Refs.~\cite{Yunes_Pretorius_09, Konno_09}. We show the scalar field for several values of $\zeta_{\rm BH}$ in Fig.~\ref{fig:theta-app}, and the metric coefficient $\omega $ in Fig.~\ref{fig:omega-app}. We see that the black hole solution exhibits the same features as the solution for a star: the scalar field first increases uniformly with $\zeta_{\rm BH}$ and then starts being damped near the horizon in the nonlinear regime $\zeta_{\rm BH} \gtrsim 1$. Frame-dragging effects become strongly screened next to the black hole horizon for large CS coupling strengths. More specifically, for $\zeta_{\rm BH} \gg 1$, screening occurs for $2M < r \lesssim 2M \zeta_{\rm BH}^{1/3}$.


\section{Constraints to the theory} \label{sec:constraints}

To lowest order in the rotation rate, the Chern-Simons modification to general relativity only affects the gravitomagnetic sector of the metric --- for large enough values of the coupling strength, however, it is possible that the prefactor of terms of order $(R\Omega)^2$ become significant in the gravitoelectric sector, but we have not addressed this issue in this paper.

Tests of the theory will therefore rely on frame-dragging effects. The CS modification changes the $t\phi$ metric component in two ways. First, it enhances the effective angular momentum $J = J_{\rm GR} + \Delta J_{\rm CS}$, defined as $J \equiv \frac12 \underset{r\rightarrow \infty}{\lim} r^3 \omega$. Secondly, it modifies the near-zone ($r \sim R$) gravitomagnetic sector around the spinning object. We will use the latter effect to set a constraint on $\ell_{\rm cs}$ with measurements of frame-dragging effects around the Earth, and the former effect to discuss potential constraints from the double-binary-pulsar system PSR J0737-3039 \cite{Burgay_03}.

\subsection{Measurements of frame-dragging around the Earth}

The Earth is clearly a nonrelativistic object, with $M_{\oplus}/R_{\oplus} \approx 7 \times 10^{-10}$. Moreover, the density of the Earth varies by no more than a factor of a few from center to edge, and the results of the analytic approximation presented in Sec.~\ref{sec:analytic-constant-density} should provide fair estimates. The change in angular momentum $\Delta J_{\rm CS}$ for an object so little compact as the Earth is completely negligible, with $\Delta J_{\rm CS}/J_{\rm GR} < \frac{4}{5} M/R < 10^{-9}$. However, the gravitomagnetic field close to the Earth's surface may be significantly modified for a large enough CS coupling constant, as one can see in Fig.~\ref{fig:earth}.

Gravity Probe B (GPB) has measured the gyroscopic precession due to frame-dragging to be in agreement with GR to an accuracy of $20\%$ \cite{GPB} . The GPB satellite is in orbit at an altitude of 640 km, which corresponds to a distance from the Earth's center $r \approx 1.1 R_{\oplus}$. This means that we have\footnote{Frame-dragging is due to the gravitomagnetic field $\vec{B}  = \vec{\nabla} \times \vec{A}$, where the gravitomagnetic potential is $\vec{A} = \frac14 r \sin \theta \omega(r) \hat{e}_{\phi}$ in spherical polar coordinates. In principle the precession of gyroscopes or the Lense-Thirring drag depend on $\omega$ and its derivative. For an approximate constraint as we give here, the detailed expression is not crucial, though, the idea being that for $\zeta \gtrsim 1$, the gravitomagnetic field becomes significantly affected.}
\beq
\Big{|}\frac{\omega(1.1 R_{\oplus})}{2 J/(1.1 R_{\oplus})^3} - 1\Big{|} \lesssim 20\%. \label{eq:Earth-constraint}
\eeq
By requiring that the constraint Eq.~(\ref{eq:Earth-constraint}) be satisfied and using Eq.~(\ref{eq:omega-out-zeta}) for $\omega$ (with $D_1 = \tanh \zeta_{\oplus}$), we obtain $\zeta_{\oplus} \lesssim 1$ (we emphasize that $\zeta$ being a dimensionless constant, it depends on the mass and radius of the system considered, hence the subscript $\oplus$).
Translating this into a constraint to $\ell_{\rm cs}$, or, to use the notation of Ref.~\cite{Yunes_Pretorius_09}, the parameter $\xi \equiv 16 \pi \ell_{\rm cs}^4$, we obtain: 
\beq
\xi^{1/4} \equiv (16 \pi)^{1/4} \ell_{\rm cs} \lesssim 10^8 \ \textrm{km}. \label{eq:lageos-constraint}
\eeq
The LAGEOS and LAGEOS 2 satellites have allowed to detect the Lense-Thirring effect for bodies orbiting the Earth with a 10\% precision \cite{Lageos}. This precision test of GR has allowed to set the first constraint on nondynamical Chern-Simons gravity \cite{Smith_08} (since then improved with measurements of the precession rate in the double-binary-pulsar system \cite{Yunes_Spergel_09, Ali-Haimoud_11}). The LAGEOS satellites are at a distance $r \approx 12000$ km $\approx 2 R_{\oplus}$ from the Earth center. Using again Eq.~(\ref{eq:Earth-constraint}) with the appropriate distance and uncertainty, we obtain the constraint $\zeta_{\oplus} \lesssim 2$, which leads, up to a factor of $\sim \sqrt 2$, to the same bound as Eq.~(\ref{eq:lageos-constraint}). 

Even though these are relatively weak constraints, they have the advantage of being independent and very robust. Even with an error of 50\% on either measurement, the constraint would be degraded by less than a factor of 3. Moreover, we will explain below that these are actually the only current astrophysical constraints (to our knowledge) of dynamical CS gravity.

\subsection{On future constraints from the double-binary-pulsar}

We have seen in the previous sections that CS gravity can only \emph{decrease} frame-dragging effects near the surface of a spinning star, and only marginally enhance them in the far-field, even for an arbitrarily large coupling strength. To be able to test CS gravity, observational accuracy must therefore reach the level at which frame dragging effects can be \emph{measured} (so one can estimate deviations from GR predictions). In other words, upper bounds on frame dragging effects \emph{cannot} be used to set any constraint on the CS coupling strength (unless the bounds are in fact lower than the GR predictions). Ref.~\cite{Yunes_Pretorius_09} derived the constraint $\xi^{1/4} \lesssim 1.5 \times10^4$ km from the measurement error on the periastron precession rate in the double-binary-pulsar system, which is much larger (by two orders of magnitude) than the GR-predicted periastron precession rate due to spin-orbit coupling. This bound was derived by extrapolating the small-coupling result to large coupling strengths. In regard of the above discussion, we see that current measurements of the periastron precession rate in the double-binary pulsar in fact cannot be used to set any constraint on CS gravity.

It has been suggested that long term measurements of the binary system PSR J0737-3039 \cite{Burgay_03} would lead to a 10\% determination of the moment of inertia of pulsar A \cite{Lyne_04,Lattimer_Schutz_05}, although this would likely take another 20 years of measurements \cite{Kramer_Wex_09}. Ref.~\cite{Yunes_et_al_10} have used this prediction to derive a potential future constraint on the CS coupling strength. From Fig.~\ref{fig:DI-full}, we see that a 10\% accuracy measurement of the moment of inertia would translate in a bound $\zeta_{\rm NS} \lesssim 1$, which would lead to the constraint\footnote{This constraint is of the same order as the one derived in Ref.~\cite{Yunes_et_al_10}, even though their metric solution actually does not contain any additional angular momentum, since they have $\delta \omega \sim 1/r^6$ at large radii. However, because they used the integral formulation (\ref{eq:IGR}) for the moment of inertia, they did obtain the correct order of magnitude for the constraint.} $\xi^{1/4} \lesssim 25$ km. Note that here we have assumed that pulsar B lies in the far field of pulsar A, i.e. that the semi-major axis $a\approx 4 \times 10^5$ km is such that $(a/R)^3 \gg \zeta_{\rm NS}$. Using the constraint from Earth measurements Eq.~(\ref{eq:lageos-constraint}), we find that $\zeta_{\rm NS}$ is actually at most $\sim 10^{13} \sim (a/R)^3$ and assuming the near-field limit would therefore not improve on the terrestrial constraint anyway.

Whereas the prospect of such a strong constraint is enticing, we should keep in mind, first, that this is a very difficult measurement \cite{Kramer_Wex_09}. Furthermore, a second challenge is the inherent uncertainty in the GR-predicted moment of inertia, which varies by $\sim 20-30\%$ depending on the equation of state used (using the LS220 or Shen et al. EOSs). In fact, Ref.~\cite{Lyne_04,Lattimer_Schutz_05} initially suggested that the measurement of the moment of inertia could help narrow down the space of allowed equations of state, since the variations in moment of inertia are larger than those in radius (by a factor of $\sim 2$ since $I \propto M R^2$), which is hard to measure in any case. We therefore conclude that constraining CS gravity from measurements of the moment of inertia is likely to be very challenging, if possible at all.

As a conclusion, the bound (\ref{eq:lageos-constraint}) that we have derived from measurements of frame-dragging around the Earth is, to our knowledge, the only current astrophysical constraint to dynamical CS gravity.

\section{Conclusions} \label{sec:conclusion}

We have presented the solution to the coupled system of field equations and the scalar field Poisson equation in dynamical CS gravity, for slowly rotating stars and black holes. We have shown that the black hole solution does not describe the spacetime outside a rotating star as was assumed in previous works. For the first time, we have provided a solution valid for an arbitrary CS coupling strength and not limited to the small-coupling regime. We have provided simple analytic solutions for nonrelativistic constant-density objects, as well as numerical solutions for realistic neutron stars and slowly-rotating black holes. 

Our solution shows two key features. First, frame-dragging effects are reduced with respect to standard GR near the surface of a rotating star or black hole. We have used this effect to set a robust constraint on the CS lengthscale from measurements of the Lense-Thirring drag and gyroscopic precession around the Earth, $\xi^{1/4} \lesssim 10^8$ km. Probes of the space-time close to the horizon of spinning black holes with orbits of objects passing nearby could potentially help constrain the theory further \cite{Konno_09}. Secondly, the angular momentum of a rotating object (as perceived by observers in the far field) is enhanced. However, this enhancement is at most $\Delta J_{\rm CS}/J_{\rm GR} \sim M/R$ and is therefore difficult to detect.

In closing, let us mention some limitations of our work. Firstly, we recall that this paper only considers the slow-rotation approximation. Arbitrarily fast rotating stars or black hole solutions are significantly more complicated, even in GR, and we have not tackled this difficult problem here. Secondly, we have only studied the Chern-Simons extension to GR and have not considered any other extension of quadratic (or higher) order in the Riemann tensor. Recently, Ref.~\cite{Pani_11} have considered compact stars in Einstein-Dilaton-Gauss-Bonnet gravity, which complements the present work for quadratic extensions to GR. Finally, we have only studied stationary spacetimes, and have not performed any stability analysis which would require a full time-dependent study. It is therefore not clear that the solutions we have found are stable, and, in fact, it is likely that instabilities occur for large coupling strengths as the system solved is effectively fourth-order. It is also possible that higher-order terms in the ``exact'' theory of which the Chern-Simons extension is a truncation render the solution stable. These questions would require a significantly more thorough analysis, which we defer to future work.

\begin{acknowledgments} 
We would like to thank Evan O'Connor for kindly providing tabulated solutions to the relativistic stellar structure equations for several neutron star masses and EOSs. We also thank Chris Hirata, Dan Grin, Nico Yunes, Frans Pretorius, Tristan Smith, Adrienne Erickcek and Marc Kamionkowski for useful conversations on Chern-Simons gravity, and Sterl Phinney for helpful discussions on spin-orbit coupling. Y. A.-H. was supported by the U.S. Department of Energy (DE-FG03-92-ER40701) and the National Science Foundation (AST-0807337) while at Caltech and is currently supported by the NSF grant AST-0807444. Y. C. is supported by by NSF grants PHY-0653653 and PHY-0601459, CAREER grant PHY-0956189, and the David and Barbara Groce start-up fund at Caltech.
\end{acknowledgments}

\end{document}